\documentclass[a4paper,11pt]{article}
\usepackage{jheppub} 

\makeatletter
\gdef\@fpheader{}
\makeatother

\usepackage{physics}
\usepackage{dsfont}
\usepackage{enumerate} 
\usepackage{enumitem}
\usepackage{todonotes}
\usepackage[cmyk]{xcolor}
\usepackage{comment}
\usepackage{placeins}

\def\tr{\text{Tr}}
\def\str{\text{tr}}
\def\AdS{\text{AdS}}

\def\I{I^{(4)}}
\def\cI{\mathcal{I}}

\def\cN{\mathcal{N}}
\newcommand{\Rm}[1]{\uppercase\expandafter{\romannumeral #1}}
\def\SU{\mathrm{SU}}

\def\Spin{\mathrm{Spin}}
\def\S{\mathrm{S}}
\def\U{\mathrm{U}}
\def\SO{\mathrm{SO}}
\def\bL{\mathbb{L}}
\def\bS{\mathbb{S}}
\def\bH{\mathbb{H}}

\def\mR{\mathcal{R}}

\def\ttL{\mathtt{L}}
\def\ttR{\mathtt{R}}
\def\ttF{\mathtt{F}}

\title{\boldmath Supersymmetric Observables from Surface Defect Anomalies in 4d and 6d SCFTs}

\author[a]{Zi-Xiao Huang,}
\author[a]{Ma-Ke Yuan}
\author[a,b,c]{and Yang Zhou}

\affiliation[a]{Department of Physics and Center for Field Theory and Particle Physics, \\Fudan University, Shanghai 200433, China}
\affiliation[b]{Peng Huanwu Center for Fundamental Theory, Hefei, Anhui 230026, China}
\affiliation[c]{Peng Huanwu Center for Fundamental Theory, Xi’an 710127, China}

\emailAdd{huangzx22@m.fudan.edu.cn}
\emailAdd{mkyuan19@fudan.edu.cn}
\emailAdd{yang\_zhou@fudan.edu.cn}

\abstract{We study the contributions of half-BPS surface defects to supersymmetric Casimir energy and the universal logarithmic term of supersymmetric R\'enyi entropy in four- and six-dimensional SCFTs. Using equivariant integration of defect anomaly polynomials and its extension to supersymmetric R\'enyi backgrounds, we obtain closed expressions for these observables in terms of defect 't~Hooft anomalies. For Gukov--Witten defects preserving transverse rotations in four-dimensional $\mathcal{N} = 4$ super Yang--Mills theories, both observables depend on a single R-symmetry anomaly. For $\mathcal{N} = (0, 4)$ defects in six-dimensional $\mathcal{N} = (1, 0)$ theories, additional terms arise from the gravitational anomaly. We provide complementary checks using chiral algebras, five-dimensional localization, and holographic probe branes at large $N$, while localized chiral fermions reproduce the gravitational-anomaly contributions. The framework also recovers Wilson-surface results in six-dimensional $\mathcal{N} = (2, 0)$ theories.}

\begin{document}
\maketitle
\flushbottom
\section{Introduction}\label{sec:introduction}
Anomalies constrain protected observables even when the dynamics of a quantum field theory is strongly coupled. 
In even-dimensional superconformal field theories (SCFTs), equivariant integration of anomaly polynomials provides a prescription for the supersymmetric Casimir energy, with an extended prescription governing universal contributions to supersymmetric R\'enyi entropy~\cite{Bobev:2015kza,Yankielowicz:2017xkf}. 
Whether this structure extends to conformal defects is a nontrivial question. Defect degrees of freedom interact with the ambient theory and generally do not form an autonomous lower-dimensional CFT. Nevertheless, their localized anomalies may determine protected contributions to observables of the coupled system.

An anomaly-based description of these contributions was developed for half-BPS Wilson surfaces in six-dimensional $\mathcal{N} = (2, 0)$ SCFTs~\cite{Huang:2025eze}. This raises two questions: how far does the prescription extend beyond this example, and what changes when a surface defect carries a gravitational anomaly? We address these questions for codimension-two Gukov--Witten defects in four-dimensional $\mathcal{N} = 4$ super Yang--Mills (SYM) theory and for codimension-four defects preserving $\mathcal{N} = (0, 4)$ supersymmetry in six-dimensional $\mathcal{N} = (1,0)$ SCFTs.
The latter provide chiral examples in which the gravitational anomaly contributes explicitly to both the energy and the entropy.

Our direct input is the defect 't~Hooft anomaly polynomial. Relations to defect Weyl anomalies provide an additional interpretation when the preserved supersymmetry relates the two sets of data. 
Throughout, anomaly control means that the contributions studied here are fixed by these anomaly coefficients on a specified supersymmetric background, without requiring a solution of the full bulk-defect dynamics.
\subsection{Defect observables}\label{subsec:intro-observables}
\begin{figure}[htp]
\centering
\includegraphics[width=\columnwidth]{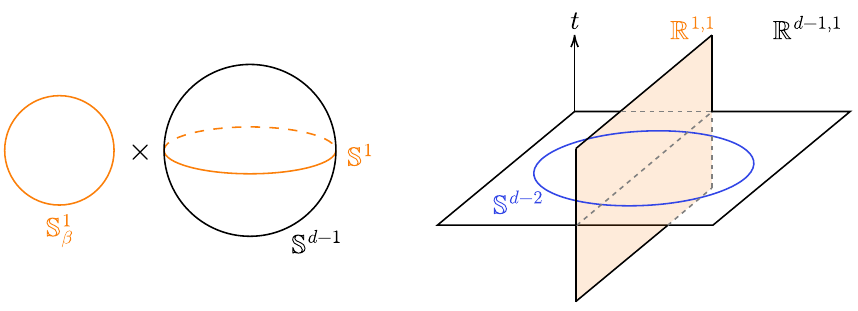} 
\caption{Geometries for the defect observables.
Left: a surface defect wrapping $\mathbb{S}^1_\beta\times \mathbb{S}^1$ in $\mathbb{S}^1_\beta\times \mathbb{S}^{d-1}$, as used to define the dSCE.
Right: a planar surface defect (orange plane) intersecting the spherical entangling surface $\mathbb{S}^{d-2}$ (blue) on the $t=0$ slice of $\mathbb{R}^{d-1,1}$, as used to define the dSRE.} 
\label{Fig:intro}
\end{figure}

We isolate the contribution of a supersymmetric surface defect $D$ through the normalized expectation value $\langle D\rangle = Z[D]/Z$, with numerator and denominator evaluated on the same background. 
For a defect wrapping $\mathbb{S}^1_\beta\times \mathbb{S}^1\subset \mathbb{S}^1_\beta\times \mathbb{S}^{d-1}$ as illustrated in Figure~\ref{Fig:intro} (Left), its contribution to the supersymmetric Casimir energy (dSCE) is defined by
\begin{equation}
  \log\langle D\rangle
  =-\beta E_C + O(1)~,
  \quad \beta\rightarrow\infty~.
  \label{eq:intro-dsce-definition}
\end{equation}
Its contribution to the supersymmetric R\'enyi entropy (dSRE) is
\begin{equation}
  S_n
  =\frac{\log\langle D\rangle_n-n\log\langle D\rangle_1}{1-n}~,
  \label{eq:intro-dsre-definition}
\end{equation}
where $n$ is the R\'enyi index and the replica background includes R-symmetry sources that preserve supersymmetry. As shown in Figure~\ref{Fig:intro} (Right), for the spherical entangling surface and planar defect considered here, the conformal map to $\mathbb{S}^1_n\times\mathbb{H}^{d-1}$ places the defect on $\mathbb{S}^1_n\times\mathbb{H}^1$ (Figure~\ref{fig-CHM})~\cite{Casini:2011kv}. 
At unit hyperbolic radius, the circle has circumference $2\pi n$. 
We focus on the universal logarithmic term in the dSRE and write $V_{\mathbb{H}^1}=2\log(\ell/\epsilon)$ for the logarithmic part of the regulated hyperbolic length, where $\ell$ and $\epsilon$ are the entangling radius and ultraviolet cutoff.

\subsection{Main results}\label{subsec:intro-results}
\paragraph{Gukov--Witten defects in four dimensions.}
We consider half-BPS Gukov--Witten surface defects in $\SU(N)$ $\mathcal{N} = 4$ SYM~\cite{Gukov:2006jk, Gukov:2008sn, Drukker:2008wr}, labeled by a Levi subgroup $\mathbb{L}=\mathrm{S}[\prod_{i=1}^{\nu}\U(N_i)]$, with $\sum_iN_i=N$. 
We restrict to $\beta_i=\gamma_i=0$, so that transverse rotations are preserved. 
Both observables are governed by the single defect R-symmetry anomaly coefficient $k_{\mathtt R}=N^2-\sum_iN_i^2$. 
The dSCE is
\begin{equation}
  E_C
  =-k_{\ttR}\frac{\Delta_2\Delta_3}{2\omega_1}~,
  \label{eq:intro-4d-dsce}
\end{equation}
where $\omega_{1,2}$ are angular chemical potentials, with $\omega_1$ associated with the defect plane, and $\Delta_{1,2,3}$ are chemical potentials for the $\SO(6)$ R-symmetry Cartans. 
They obey $\omega_1+\omega_2+\Delta_1+\Delta_2+\Delta_3=0$ in our conventions.
For the replica backgrounds studied here, only the $R_2$ and $R_3$ sources vary with $n$, with weights $s_2+s_3=1$. We obtain
\begin{equation}
  S_n
  =\frac{k_{\mathtt R}}{2}
   \left[s_2 s_3\Big(\frac1n-1\Big) + 1\right]V_{\mathbb H^1}~.
  \label{eq:intro-4d-dsre}
\end{equation}
For these defects, $b=d_2=3k_{\mathtt R}$, so the results also admit a description in terms of defect Weyl-anomaly coefficients.
\paragraph{Chiral surface defects in six dimensions.}
For a half-BPS surface defect preserving $\mathcal N=(0,4)$ supersymmetry in a six-dimensional $\mathcal N=(1,0)$ SCFT, with additional flavor backgrounds set to zero, we use the defect anomaly polynomial
\begin{equation}
  I^{(4)}
  =\frac{k_L}{2}c_2(L)+\frac{k_R}{2}c_2(R)
   +\frac{k_I}{2}c_2(I)-\frac{k_g}{24}p_1(T)~.
  \label{eq:intro-6d-anomaly}
\end{equation}
Here $\SU(2)_L\times\SU(2)_R$ is the double cover of the transverse rotation group, $SU(2)_I$ is the ambient R-symmetry, and $k_g=c_L-c_R$ denotes the gravitational anomaly. 
Equivariant integration gives the dSCE
\begin{equation}
  E_C
  =-\frac{k_L(\omega_2+\omega_3)^2
          +k_R(\omega_2-\omega_3)^2+k_I\sigma^2}{8\omega_1}
   -\frac{k_g\omega_1}{24}~,
  \label{eq:intro-6d-dsce}
\end{equation}
where $\omega_{1,2,3}$ are angular chemical potentials, the defect lies in the $\omega_1$ plane, and the $\SU(2)_I$ chemical potential satisfies $\sigma=\omega_1+\omega_2+\omega_3$. 
On the round supersymmetric replica background considered here, with replica dependence only in the $\SU(2)_I$ source, the dSRE becomes
\begin{equation}
  S_n
  =\left[
     \frac{k_I}{8}\Big(\frac{1}{n} + 7\Big)
     +\frac{k_g}{24}\Big(\frac{1}{n} + 1\Big)
   \right]V_{\mathbb H^1}~.
  \label{eq:intro-6d-dsre}
\end{equation}
This form separates the R-symmetry and gravitational-anomaly contributions.
Although $k_L$ and $k_R$ enter the dSCE, they drop out of the dSRE on this background. In both dimensions, the logarithmic entropy coefficient has the simple form $A/n+B$, with $A$ and $B$ fixed by the relevant anomaly data and background parameters.

\subsection{Strategy and complementary checks}
\label{subsec:intro-strategy}
We adopt the defect analogue of the anomaly-polynomial prescription,
\begin{equation}
  E_C=\int_{\mathbb{R}^2} I^{(4)}~,
  \label{eq:intro-equivariant-prescription}
\end{equation}
where the integral is equivariant with respect to the rotations and R-symmetry transformations preserved by the defect. 
The dSRE follows from an extended equivariant prescription with an auxiliary rotation parameter encoding the replica index~\cite{Yankielowicz:2017xkf,Huang:2025eze}. 
We test these prescriptions through complementary descriptions of the defects.

In four dimensions, a two-dimensional gauged linear sigma model and anomaly inflow yield the same Gukov--Witten defect anomaly polynomial. 
The chiral-algebra description checks the dSCE in the Schur limit, while a probe D3-brane calculation reproduces the leading large-$N$ dSRE for the simple Levi subgroup $\mathrm{S}[\U(1)\times \U(N-1)]$.

In six dimensions, we study defects preserving large $\mathcal N=(0,4)$ supersymmetry, engineered by D4-D2 branes in type IIA constructions of $\mathcal{N} = (1,0)$ SCFTs~\cite{Apruzzi:2024ark}. 
For the massless type IIA, equal-rank family, localization in the five-dimensional $\mathcal N=1$ circular quiver obtained by compactification checks the $k_I$ contribution to the dSCE in the large-$N$ probe regime. 
In the same family, the leading $k_I$ contribution to the dSRE is independently reproduced by a probe D4-brane in the uplift of a charged hyperbolic black hole. 
Determinants of localized chiral fermions account for the $k_g$ terms in both observables; these gravitational-anomaly terms are not captured by the classical probe calculation. We also apply the prescription to the known $(2,0)$ Wilson-surface dSCE,
together with the dSRE on the corresponding equal-weight R-symmetry background~\cite{Huang:2025eze}.

The paper is organized as follows. Section~\ref{sec:4d} studies Gukov--Witten defects and their anomaly, chiral-algebra, and holographic descriptions. 
Section~\ref{sec:6d} develops the six-dimensional results and their applications to D4-D2 defects. Section~\ref{sec:discussion} discusses open questions. 
Technical details are collected in the appendices.
\section{\boldmath Gukov--Witten defect in 4d $\cN=4$ SYM}\label{sec:4d}
\subsection{Review of Gukov--Witten surface defect}\label{subsec:rev4d}
We begin with a brief review of half-BPS Gukov--Witten (GW) surface defects in 4d $\cN=4$ $\SU(N)$ SYM~\cite{Gukov:2006jk, Gukov:2008sn, Drukker:2008wr}. 
The GW defect is a codimension-two surface operator supported on a 2d submanifold $\Sigma$ inside the 4d ambient spacetime.
It is defined by prescribing singular behavior near $\Sigma$ for the 4d $\cN=4$ gauge field $A$ and the complex scalar $\Phi=\frac{1}{\sqrt{2}}(\phi^5+i\phi^6)$. 
In local coordinates
$z=re^{i\theta}$ transverse to $\Sigma$, the singular profiles take the
form
\begin{equation}
A=
\begin{pmatrix}
\alpha_{1}\otimes 1_{N_1}&0 &\cdots & 0\cr
0& \alpha_{2}\otimes1_{N_2}&\cdots & 0 \cr
\vdots &\vdots &\ddots&\vdots\cr
0&0&\cdots&\alpha_{\nu}\otimes1_{N_\nu}
\end{pmatrix}
\dd \theta~,
\end{equation}
as well as
\begin{equation}
\label{eq:scalar-sing}
\Phi=\frac{1}{\sqrt{2}z}
\begin{pmatrix}
(\beta_{1}+i\gamma_1)\otimes 1_{N_1}&0 &\cdots & 0\cr
0& (\beta_{2}+i\gamma_2)\otimes1_{N_2}&\cdots & 0 \cr
\vdots &\vdots &\ddots&\vdots\cr
0&0&\cdots&(\beta_{\nu}+i\gamma_\nu)\otimes1_{N_\nu}
\end{pmatrix}~.
\end{equation}
The block structure of the singularity specifies a Levi
subgroup $\mathbb{L}$ labeled by the partition $[N_i]$ of $N$,
\[
\bL=\S\bigg[\prod_{i=1}^{\nu}\U(N_i)\bigg] \subset \SU(N)~, \quad  \sum_{i=1}^{\nu}N_i=N~.
\]
In addition to the singularity of $A$ and $\Phi$, the surface operator can carry a two-dimensional theta angle parameter,
\begin{equation}
\eta=
    \begin{pmatrix}
     \eta_{1}\otimes 1_{N_1}&0 &\cdots & 0\cr
0& \eta_{2}\otimes1_{N_2}&\cdots & 0 \cr
\vdots &\vdots &\ddots&\vdots\cr
0&0&\cdots&\eta_{\nu}\otimes1_{N_\nu}
\end{pmatrix}~,
\end{equation}
which enters the action through a coupling term $\sum_{i = 1}^\nu\eta_i \int_\Sigma\Tr F^{(i)}$, where $F^{(i)}$ is the field strength of the unbroken $\U(N_i)$ factor.
The GW surface defect is specified by the choice of Levi subgroup $\bL$, together with the continuous parameters $(\alpha,\beta,\gamma,\eta)$.

Specifically, introducing a singular behavior~\eqref{eq:scalar-sing} for the complex scalar field $\Phi$ breaks the rotation symmetry in directions orthogonal to the defect. In this work we consider the GW defect that preserves the transverse rotation symmetry, with moduli parameters $\beta_i=\gamma_i=0$.

\paragraph{The symmetry algebra.}
For later convenience, we summarize the symmetry algebra preserved by the GW defect.
The maximal bosonic subalgebra of the 4d $\cN=4$ superconformal algebra is
\begin{equation}
\mathfrak{so}(4,2) \times \mathfrak{su}(4)_R \subset \mathfrak{psu}(2,2|4)~,
\end{equation}
where $\mathfrak{so}(4,2)$ is the 4d conformal algebra and $\mathfrak{su}(4)_R$ is the R-symmetry algebra.
A half-BPS GW defect preserves a 2d small $(4,4)$ superconformal symmetry on the defect submanifold $\Sigma$, with
the superconformal algebra~\cite{Drukker:2008wr, Bomans:2024vii}
\begin{equation}
\qty(\mathfrak{psu}(1,1|2)\times \mathfrak{psu}(1,1|2)) \ltimes \mathfrak{u}(1)_C \subset \mathfrak{psu}(2,2|4)~, 
\end{equation}
where $\mathfrak{psu}(1,1|2)\times\mathfrak{psu}(1,1|2)$ is the global part of the small 2d $\cN=(4,4)$ superconformal algebra on the defect, while $\mathfrak{u}(1)_C$ is its central extension, and it is generated by a diagonal combination of the rotation $\mathfrak{so}(2)_\perp$ transverse to the defect and a $\mathfrak{u}(1)_\ttR\subset\mathfrak{su}(4)_R$ R-symmetry.

In the following part of this paper, we restrict to GW defects with $\beta_i=\gamma_i=0$. 
In this case, the transverse rotation symmetry $\mathfrak{so}(2)_\perp$ is restored, and an additional $\mathfrak{u}(1)_A$ symmetry, which acts as an outer automorphism of the small $\cN=(4,4)$ superconformal algebra, is preserved.
The symmetry algebra relevant for our discussion is therefore
\begin{equation}
\mathfrak{u}(1)_A
\ltimes
\qty(
\mathfrak{psu}(1,1|2)\times\mathfrak{psu}(1,1|2)
)
\ltimes
\mathfrak{u}(1)_C
\subset
\mathfrak{psu}(2,2|4)~.
\end{equation}

The two $\mathfrak{psu}(1,1|2)$ factors contain the $\mathfrak{su}(2)_\ttL\times\mathfrak{su}(2)_\ttR \subset\mathfrak{su}(4)_R$ R-symmetry of the 2d $\cN=(4,4)$ theory.\footnote{In Section~\ref{sec:4d}, we use typewriter-font labels $\ttL$ and $\ttR$ for the $\SU(2)_{\ttL,\ttR}$ R-symmetries preserved by the defect, to distinguish them from quantities associated with the 4d $\cN = 2$ R-symmetry $\SU(2)_R\times \U(1)_r$. We also use $\ttF$ to denote the defect flavor symmetry. }
Without loss of generality, we consider the surface defect inserted in the $x^{0,1}$-plane. The 2d Cartan generators are then identified with the 4d ones as follows~\cite[Appendix B]{Chen:2023lzq},\footnote{We use a Cartan normalization differing from that of~\cite{Chen:2023lzq} by $R_I^{\rm here}=\frac{1}{2}R_I^{\rm there}$, such that the $\mathbf4$ of $\SU(4)_R$ carries weights $\pm \frac{1}{2}$.}
\begin{equation}\label{eq-cartan-id1}
\begin{aligned}
&L_0= \frac{E+J_1}{2}~, \quad \bar{L}_0 = \frac{E-J_1}{2}~, \quad J_\ttL =\frac{R_2-R_3}{2}~, \quad J_\ttR =\frac{R_2+R_3}{2}~, \\
&J_A = R_1-R_2~,\quad J_C=-J_2-R_1~,
\end{aligned}
\end{equation}
where $L_0$ and $\bar L_0$ are the 2d left- and right-moving Hamiltonians, $J_\ttL$ and $J_\ttR$ are the Cartan generators of the 2d left- and right-moving $\mathfrak{su}(2)$ R-symmetries, $J_A$ generates the outer-automorphism $\mathfrak{u}(1)_A$, $J_C$ generates the central extension $\mathfrak{u}(1)_C$, $J_1, J_2$ are the Cartan generators of the 4d rotation symmetry $\mathfrak{so}(4) \subset \mathfrak{so}(4,2)$, and $R_1, R_2, R_3$ are the Cartan generators of the $\mathfrak{su}(4)_R$ R-symmetry.

\begin{figure}[t] 
\centering
\includegraphics[width=0.8\textwidth]{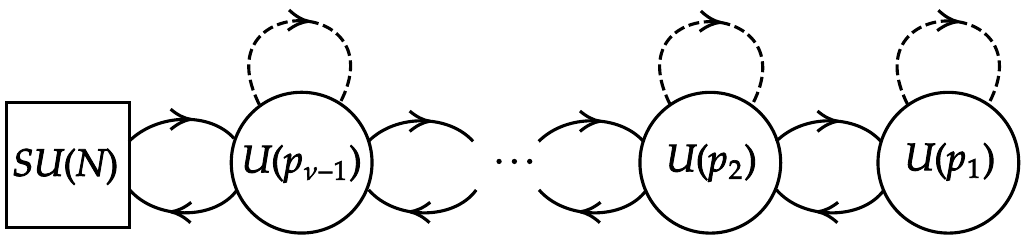} 
\caption{The linear quiver diagram of the 2d GLSM description corresponding to the GW surface defect. We adapt the $(2,2)$ field content in the figure. The dashed line represents the $(2,2)$ chiral multiplet $X_i$ for each gauge group $\U(p_i)$, $i=1,\dots,\nu-1$. There is also one $(2,2)$ vector multiplet $V_i$ for each gauge group $\U(p_i)$. The solid lines connecting $\U(p_i)$ and $\U(p_{i+1})$ represent the $(2,2)$ (anti-)bifundamental chiral multiplets $\Phi_{i(i+1)}$ and $\Phi_{(i+1)i}$. The 2d GLSM is coupled to the 4d ambient theory by gauging the global symmetry $\SU(N)$.} 
\label{Fig:2d-GLSM}
\end{figure}

\paragraph{The 2d GLSM description.}
An alternative description of the GW defect is to couple a 2d $\cN = (4,4)$ gauged linear sigma model (GLSM) to the 4d $\cN=4$ SYM theory~\cite{Gadde:2013dda}. The 2d GLSM lives on the defect plane $\Sigma$ and serves as the UV description of the surface defect.  
For the GW defect labeled by the Levi group $\bL=\S[\prod_{i=1}^{\nu}\U(N_i)]$, the corresponding GLSM is a linear quiver theory
\[
\qty[\SU(p_\nu)] \times \U(p_{\nu-1}) \times \cdots \times \U(p_{1})
\]
with $p_i = \sum_{j=1}^i N_j$ and $p_\nu=N$. 
The 2d GLSM is coupled to the 4d $\SU(N)$ $\cN=4$ SYM ambient theory by gauging the 2d flavor symmetry $\SU(N)$. The 2d GLSM field content consists of one $(4,4)$ vector multiplet for each gauge node $\U(p_{i})$, and one $(4,4)$ bifundamental hyper multiplet between each two adjacent nodes; see Figure~\ref{Fig:2d-GLSM} and also Appendix~\ref{apdx:GW-field}.
In particular, for the \emph{simple} defect labelled by $\bL=\S[\U(1)\times \U(N - 1)]$ with the partition
\[
N_i=[\underbrace{0,\cdots,0}_{\nu - 2},1,N-1]~,
\]
the corresponding 2d GLSM is a $\U(1)$ gauge theory with $\SU(N)$ flavor symmetry, i.e., a 2d $\cN=(4,4)$ supersymmetric QED with $N$ hypermultiplets.

\paragraph{Defect anomaly data.}
The defect Weyl-anomaly coefficient $b$ can be extracted from the IR central charge of the 2d $(4,4)$ GLSM description. 
As a UV description of the GW defect, the 2d $(4,4)$ GLSM flows in the infrared (IR) to a nonlinear sigma model (NLSM) with target space $T^*(G/\bL)$. 
Through $b$-extremization mechanism~\cite{Wang:2020xkc}, the defect Weyl anomaly $b$ is identified with the IR central charge $c_{2d}$. With the corresponding IR $\U(1)_\ttR$ R-symmetry being the Cartan subgroup of $\SU(2)_\ttR$, the $b$-anomaly is given by
\begin{equation}
b = c_{2d} = 3k_\ttR = 3\bigg(N^2 - \sum_{i = 1}^\nu N_i^2\bigg)~,
\end{equation}
where $k_\ttR$ is the 't~Hooft anomaly of $\SU(2)_\ttR$.\footnote{We normalize the $\SU(2)_\ttR$ charge to be $\pm \frac{1}{2}$, and the $\U(1)_\ttR$ charge to be $\pm 1$, i.e., the generator of $\U(1)_\ttR$ is $2J_\ttR$. We also define $k_\ttR$ to be twice the $\SU(2)_\ttR$ current algebra level $\kappa_\ttR$. Therefore, in our convention we have $b=c_{2d}=6\kappa_\ttR=3k_\ttR$.}
Equivalently, the $b$-anomaly can also be read off from the NLSM description, as the complex dimension of $T^*(G/\bL)$ fixes the IR central charge~\cite{Gadde:2013dda},
\begin{equation}
b=c_{2d}=3 \dim_\mathbb{C} T^*(G/\bL)=3\bigg(N^2-\sum_{i=1}^\nu N_i^2\bigg)~.
\end{equation}
\subsection{Defect in the chiral-algebra limit}
\label{subsec: defect-chiral}
In this subsection we consider the supersymmetric partition function of a 4d $\mathcal{N}=4$ $\SU(N)$ SYM on $\mathbb{S}^1_\beta\times\mathbb{S}^3_{\vec\omega}$ with a single $\frac{1}{2}$-BPS GW defect inserted on $\mathbb{S}^1_\beta \times \mathbb{S}^1_{\omega_\ell}$. 
The GW defect is labeled by the Levi group $\bL$ and we consider the defect parameters $\beta_i=\gamma_i=0$.
We also consider a simple but interesting limit of the supersymmetric partition function $Z$, called the Schur limit, where  dramatic simplifications appear due to a particular setting of the fugacities, relating $Z$ to the vacuum character of the associated 2d chiral algebra sector.\footnote{We will use the terminology of Schur/chiral-algebra limit interchangeably.}
In the Schur/chiral-algebra limit,
the defect expectation value is computed in~\cite{Chalabi:2020iie}, from which we extract the chiral-limit dSCE~\eqref{eq:dsce-schur} and show that it is proportional to the defect 't~Hooft anomaly~$k_\ttR$.

Four-dimensional SCFTs with $\cN\geq2$ supersymmetry admit a protected sector of operators and algebras isomorphic to a two-dimensional chiral algebra~\cite{Beem:2013sza}.
The chiral algebra sector is defined by passing to the cohomology of a certain nilpotent supercharge $Q$, restricting the associated local operators to lie on a fixed two-dimensional plane (the chiral-algebra plane) $\mathbb{R}^2 \subset \mathbb{R}^4$.

The 4d $\cN=2$ supersymmetric partition function $Z$ is given by the superconformal index $\cI$ combined with the supersymmetric Casimir energy $\mathsf{E}_C$,
\begin{equation}
    Z=e^{-\beta \mathsf{E}_C}\cI~.
\end{equation}
The canonical definition of the superconformal index $\cI$ is~\cite{Bobev:2015kza}
\begin{equation}
\label{eq:4d-N=2-index}
    \cI_{\cN=2} = \tr \qty[ (-1)^F p^{J_1-r}q^{J_2-r}t^{r+R}]~,
\end{equation}
where $p\equiv e^{-\beta \omega_1},q\equiv e^{-\beta \omega_2},t\equiv e^{-\beta \mu}$ are the fugacities, $J_{1,2}$ are the Cartans of the $\SO(4)$ rotation symmetry, $R, r$ are the Cartans of the $\SU(2)_R\times\U(1)_r$ R-symmetry. The superconformal index~\eqref{eq:4d-N=2-index} preserves the supercharge $Q$ with eigenvalues $(E,J_1,J_2,R,r)=(\frac{1}{2},-\frac{1}{2},-\frac{1}{2},\frac{1}{2},-\frac{1}{2})$ and the commutator $\acomm{Q}{Q^\dagger}=E-J_1-J_2-2R+r$.
The SCE $\mathsf{E}_C$ is universally determined by the 4d conformal anomaly $a, c$ and the fugacities $p,q,t$~\cite{Bobev:2015kza}.

It is shown in~\cite{Beem:2013sza,Bobev:2015kza} that the Schur limit $q=t$ relates the supersymmetric partition function $Z$ to the character of the vacuum representation $V_0$ of the 2d chiral algebra sector associated with the supercharge $Q$ and the chiral plane labeled by $\omega_2$,
\begin{equation}
    Z_0=\tr_{V_0}\qty(q^{\mathsf{L}_0-\frac{c_{0}}{24}})~,
\end{equation}
where $\mathsf{L}_0$ is the zero mode of the 2d chiral algebra stress tensor, $c_0=-12c$ is the 2d central charge coming from the contribution of the SCE,
\begin{equation}
    e^{-\beta \mathsf{E}_C}|_{\text{chiral limit}}=q^{c/2}~.
\end{equation}
Notice that $Z$ only depends on $q$ in the Schur limit.

The chiral-algebra limit continues to be well defined in the presence of a surface defect preserving at least $\mathcal N=(2,2)$ supersymmetry, provided that the defect is inserted orthogonally to the chiral plane~\cite{Cordova:2017mhb}. In this configuration, the defect preserves the nilpotent supercharge $Q$ that defines the chiral-algebra cohomology. We may therefore consistently study defect insertions within the chiral-algebra limit.
The defect insertion leads to a non-vacuum module of the 2d chiral algebra~\cite{Cordova:2017mhb,Bianchi:2019sxz}.

Under the chiral-algebra limit $q=t$, a GW defect labeled by the Levi group $\bL$ in a 4d $\mathcal{N}=4$ theory corresponds to a local operator $O_\mR$ on $\bS^1$ in 2d $q$-deformed $\SU(N)$ YM theory on $\mathbb{S}^1\times \bS^1$, with the representation $\mR$ specified by $\bL$~\cite{Alday:2013kda}. 
The defect expectation value is given by the one-point function of $O_\mR$,
\begin{equation}
\label{eq:2d-qYM}
   \expval{O_\mR} = q^{-\frac{1}{2}\qty(N^2-\sum_{i=1}^\nu N_i^2)}\qty(1+{\cal O}(q^\delta))~, \quad \delta > 0~.
\end{equation}
To extract the dSCE, the $q$-exponent factor needs further careful normalization.
When the surface defect intersects the chiral-algebra plane orthogonally, the defect gives rise to a non-vacuum module $V_\sigma$ of the 2d chiral algebra.
The surface defect itself, with no additional local operator insertions on the defect, is seen by the 2d chiral algebra as a point operator $\sigma(0)$ inserted at the origin of the 2d chiral plane, which is called the defect identity in the sense of the $Q$-cohomology class. This operator generates a non-vacuum state $\ket{\sigma}\in V_\sigma$ with conformal weight $h_\sigma$.
For the GW defect in 4d $\cN=4$ $\SU(N)$ SYM, 
\begin{equation}
    h_\sigma=-\frac{1}{4}\bigg(N^2-\sum_{i=1}^\nu N_i^2\bigg)~.
\end{equation}
This leads to a $q^{h_\sigma}$ factor in the Schur-limit of the defect expectation value~\cite{Bianchi:2019sxz},
\begin{equation}
   \expval{D} = \frac{Z_D}{Z_0}=\frac{\Tr_{V_\sigma} q^{\mathsf{L}_0-\frac{c_0}{24}}}{\Tr_{V_0} q^{\mathsf{L}_0-\frac{c_0}{24}}}=q^{h_\sigma}\,\qty(1+{\cal O}(q^\delta))~, \quad \delta > 0~,
\end{equation}
with the structure $(1+{\cal O}(q^\delta))$ fixed by~\eqref{eq:2d-qYM}.
Thus, taking the large $\beta$ limit, we extract the dSCE in the chiral-algebra limit,
\begin{equation}
\label{eq:dsce-schur}
 E_C|_{\text{chiral limit}}=-\frac{1}{4}\bigg(N^2-\sum_{i=1}^\nu N_i^2\bigg)\omega_2=-\frac{1}{4}\,k_\ttR\,\omega_2~.
\end{equation}

\subsection{Defect anomaly polynomial}\label{subsec:GW-defect-ap}
In this subsection we compute the anomaly polynomial of the GW defect from both the 2d GLSM description and the inflow mechanism. We then obtain the closed formula of dSCE and dSRE from the equivariant integration of the defect anomaly polynomial, showing that these two quantities are controlled by the defect 't~Hooft anomaly $k_\ttR$ of the $\SU(2)_\ttR$ R-symmetry preserved by the defect.
\subsubsection{Derivation from the 2d GLSM and inflow mechanism}
In the following, we derive the anomaly polynomial of the GW defect in two complementary ways. We first compute it from the two-dimensional GLSM description, and then reproduce the same result using the anomaly inflow mechanism. The two derivations provide a direct consistency check of the defect anomaly polynomial~\eqref{eq:GW-ap-general}. Details of the derivation from the GLSM description are collected in Appendix~\ref{appx:GW-anomaly-polynomial}.
\paragraph{From 2d GLSM description.}

For the general case $\bL=\S[\prod_{i=1}^{\nu}\U(N_i)]$, the 2d $(4,4)$ GLSM is a linear quiver theory $[\SU(p_\nu)]\times\U(p_{\nu-1})\times \cdots \U(p_{1})$. The field content consists of one 2d $(4,4)$ vector multiplet for each gauge node, and one 2d $(4,4)$ bifundamental hypermultiplet associated to each two adjacent nodes,
as described in Section~\ref{subsec:rev4d}. Denote the fermions from the $(4,4)$ vector multiplets and hypermultiplet as $\lambda_{\pm},\psi_{\pm}$, respectively.
They transform under the global symmetry groups $\SU(2)_\ttL\times\SU(2)_\ttR\times\SU(2)_\ttF$ as\footnote{We follow the convention in~\cite{Wang:2020xkc}.}  
\begin{equation}
\label{eq:GLSM-charge}
\begin{aligned}
    (\lambda_+,\lambda_-):(\bold{2},\bold{1},\bold{2})_+\oplus (\bold{1},\bold{2},\bold{2})_-~,\\
    (\psi_+,\psi_-):(\bold{1},\bold{2},\bold{1})_+\oplus (\bold{2},\bold{1},\bold{1})_-~,
\end{aligned}
\end{equation}
depending on chirality. The Cartan generator of $\SU(2)_\ttF$ is $R_1$.
The two fermions of opposite chirality in each supermultiplet have identical contributions to the gauge anomaly, the gravitational anomaly,
and the 't~Hooft anomaly of the $\SU(2)_\ttF$ global symmetry. These anomaly terms therefore cancel, leaving only the anomaly terms of the $\SU(2)_\ttL\times \SU(2)_\ttR$ R-symmetry.
At the $i$-th gauge node, the vector multiplet contributes to the defect anomaly polynomial as
\begin{equation}
    I^{(i)}_{\rm vec}
    =
    \frac{p_i^2}{2}
    \left(
        \str_{\rm fund}F_\ttL^2
        -
        \str_{\rm fund}F_\ttR^2
    \right),
\end{equation}
while the bifundamental hypermultiplet contributes
\begin{equation}
    I^{(i)}_{\rm hyp}
    =
    -\frac{p_i p_{i+1}}{2}
    \left(
        \str_{\rm fund}F_\ttL^2
        -
        \str_{\rm fund}F_\ttR^2
    \right).
\end{equation}
$F_{L,R}$ are the background field strength associated with $\SU(2)_{\ttL,\ttR}$, and $c_2(F)=\frac{1}{2}\str_{\text{fund}}F^2$ denotes the second Chern class.
Summing up the contribution of fermions associated with each node, the defect anomaly polynomial is given by
\begin{equation}
\label{eq:GW-ap-general}
\begin{split}
\I &=-\frac{1}{2}\sum_{i=1}^{\nu-1}(p_ip_{i+1}-p_i^2)(\str_{\text{fund}}F_\ttL^2-\str_{\text{fund}}F_\ttR^2)\\
&=-\frac{1}{2}\qty\Big(N^2-\sum_{i=1}^{\nu}N_i^2)(c_2(\ttL)-c_2(\ttR))~,
\end{split}
\end{equation}
where we have used the relation $p_i = \sum_{j=1}^i N_j$ and $p_\nu=N$.

\paragraph{From the inflow mechanism.}
The defect anomaly polynomial can also be obtained independently from the anomaly inflow mechanism. We extend the type IIB inflow formalism of 4d $\cN=4$ SYM in~\cite{Bah:2020jas} to the bulk-to-defect geometry in~\cite{Bomans:2024vii}.\footnote{The result quoted in~\cite{Bomans:2024vii} contains typos and does not implement the normalization in which the trivial defect has vanishing defect anomaly. We correct both issues below.}

Consider the general case of the GW defect, with the partition of $N$ as $[w_1^{k_1},\dots,w_p^{k_p}]$,
\begin{equation}
N \to \big(\underbrace{w_1,\dots, w_1}_{k_1},\underbrace{w_2,\dots, w_2}_{k_2},\dots,\underbrace{w_p,\dots, w_p}_{k_p}\big)~, \qquad \sum_{a=1}^{p}w_a k_a=N~.
\end{equation}
For later convenience, it can be written in the previous notation as
\begin{equation}
    N \to \qty(N_1, N_2, \dots, N_\nu)~,  \qquad
    \nu=\sum_{a=1}^{p}k_a ,
\end{equation}
where each $w_a$ appears $k_a$ times.

In the type IIB description, the bulk-defect system is realized in a warped $\AdS_3\times Y_7$ geometry, where the $\AdS_3$ factor extends along the surface defect, and $Y_7$ is a non-compact
seven-manifold transverse to it. Away from the defect, the geometry approaches $\AdS_5\times S^5$. Near the defect, the geometry is described by a local puncture geometry $X_7$, labelled by the partition data $[w_1^{k_1},\ldots,w_p^{k_p}]$ that specifies the GW defect.

For anomaly inflow, one introduces the auxiliary eleven-dimensional space $Y_{11}$, obtained by fibering $Y_7$ over a four-manifold $X_4$. The non-compact $Y_7$ can be separated into the asymptotic bulk region, a cylindrical interpolation region, and the local puncture geometry $X_7$. The asymptotic region gives the anomaly of the four-dimensional bulk theory, while the cylindrical region does not contribute. Consequently, the contribution localized on the defect is precisely the inflow from the local puncture geometry,
\begin{equation}
    \I_{\text{punc}}(w_a,k_a)=-\frac{1}{2}\bigg(\sum_{a=1}^{p}l_a(w_a^2-w_{a-1}^2)\bigg)\chi(\SO(4)_{\perp})~,
\end{equation}
where $l_a=\sum_{b=a}^{p}k_b$, $w_0\equiv0$.
We subtract the puncture contribution associated with the trivial partition $[N]$,
\begin{equation}
    \I_{\text{punc}}([N])=-\frac{1}{2}N^2\chi(\SO(4)_{\perp})~,
\end{equation}
so that the trivial defect has vanishing anomaly.
The resulting defect-localized inflow is
\begin{equation}
\label{eq:GW-ap-inflow}
\I_{\text{inflow}} = \I_{\text{punc}}(w_a,k_a) - \I_{\text{punc}}\qty([N]) = \frac{1}{2}\bigg(N^2-\sum_{a=1}^{p}l_a(w_a^2-w_{a-1}^2)\bigg)\chi(\SO(4)_{\perp})~.
\end{equation}
The defect anomaly polynomial for a GW defect with generic partition $[w_1^{k_1},\dots,w_p^{k_p}]$ is
\begin{equation}
\label{eq:inflow-ap}
    \I_{\text{def}}=-\I_{\text{inflow}}=-\frac{1}{2}\bigg(N^2-\sum_{a=1}^{p}l_a(w_a^2-w_{a-1}^2)\bigg)\chi(\SO(4)_{\perp})~.
\end{equation}
To make the agreement with the field-content computation manifest, we rewrite the anomaly coefficient as
\begin{equation}\label{eq:kR-inflow}
\begin{aligned}
     k_\ttR&\equiv N^2-\sum_{a=1}^{p}l_a(w_a^2-w_{a-1}^2)
     =N^2-\sum_{a=1}^{p} w_a^2 k_a
     =N^2-\sum_{i=1}^{\nu}N_i^2~.
\end{aligned}
\end{equation}
Using also $\chi(\SO(4)_{\perp})=c_2(\ttL)-c_2(\ttR)$,\footnote{See Appendix~\ref{appx:conv} for our conventions for the Euler class and the second Chern classes.} 
the inflow result~\eqref{eq:inflow-ap} agrees exactly with the field-theory result~\eqref{eq:GW-ap-general}.
\subsubsection{Equivariant integral}
\label{sec:dSRE-int-4d}
We now extract the dSCE and dSRE from the GW defect anomaly polynomial~\eqref{eq:GW-ap-general} obtained above. We first compute the equivariant integral of the defect anomaly polynomial to obtain the dSCE~\eqref{eq:GW-dSCE-kR}.
We then derive the dSRE~\eqref{eq:GW-dSRE-2} from extending the equivariant integral to the replica geometry with the allowed R-symmetry background preserved by the GW defect insertion. These two observables are both entirely controlled by the GW defect 't~Hooft anomaly $k_\ttR$.
\paragraph{dSCE.}
For the dSCE geometry, consider the $\frac{1}{2}$-BPS GW defect inserted on $\mathbb{S}^1_\beta\times\mathbb{S}^1_{\omega_\ell} \subset \mathbb{S}^1_\beta\times\mathbb{S}^3_{\vec\omega}$. 
Without loss of generality, we choose that the defect is inserted along the $\omega_1$-plane, i.e. $\ell =1$. 
The defect preserves a two-dimensional small $(4,4)$ superconformal algebra with  R-symmetry $\SU(2)_\ttL\times \SU(2)_\ttR$, whose Cartan generators are identified as in~\eqref{eq-cartan-id1}. In the UV GLSM description, there is in addition an $\SU(2)_\ttF$ flavor symmetry, whose Cartan generator is identified as $J_\ttF=R_1$.\footnote{See Appendix C of~\cite{Chen:2023lzq}.}

The equivariant integral of~\eqref{eq:GW-ap-general} gives
\begin{equation}\label{eq-dSCE-general}
\int_{\mathbb{R}^2_{\omega_1}} \I =
\frac{1}{8\omega_1}\bigg(N^2-\sum_{i=1}^{\nu}N_i^2\bigg)\qty(a_\ttL^2-a_\ttR^2)~,
\end{equation}
where $a_{\mathtt{L},\mathtt{R}}$ are the chemical potentials corresponding to the $\SU(2)_{\ttL,\ttR}$ R-symmetry preserved by the defect.
To express the defect equivariant parameters in terms of the
four-dimensional variables, we parameterize the supersymmetric background
by the angular chemical potentials $\omega_i$ and the $\SU(4)_R$ Cartan
chemical potentials $\Delta_I$, which enter the $\cN=4$ superconformal
index as
\begin{equation}
    I_{\cN=4}
    =
    \tr\qty[
    (-1)^F
    e^{-\beta\left(
    \sum_{i=1}^2\omega_iJ_i
    +
    \sum_{I=1}^3\Delta_I R_I
    \right)}
    ]~,
\end{equation}
subject to the BPS constraint
\begin{equation}
    \sum_{i=1}^2\omega_i+\sum_{I=1}^3\Delta_I=0~.
\end{equation}
Using the Cartan generators identification~\eqref{eq-cartan-id1} as well as the pullback
\begin{equation}
a_\ttL J_\ttL + a_\ttR J_\ttR + a_\ttF J_\ttF = \sum_{i=1}^3 \Delta_i R_i~,
\end{equation}
the chemical potentials are identified as
\begin{equation}
a_\ttL = \Delta_2 - \Delta_3~, \quad a_\ttR = \Delta_2 + \Delta_3~, \quad a_\ttF = \Delta_1~.
\end{equation}
Hence~\eqref{eq-dSCE-general} can be expressed by the 4d bulk chemical potentials as
\begin{equation}
\label{eq-equiv-int}
E_C = \int_{\mathbb{R}^2_{\omega_1}} \I = -\frac{1}{2}\bigg(N^2-\sum_{i=1}^{\nu}N_i^2\bigg)\frac{\Delta_2 \Delta_3}{\omega_1}~.
\end{equation}
Now we examine the chiral-algebra limit. As in Section~\ref{subsec: defect-chiral}, we choose the $\omega_2$-plane to be the chiral plane. The defect insertion is thus orthogonal to the chiral plane and preserves the chiral-algebra limit.
Under the Schur limit setting\footnote{See Appendix~\ref{app:Schur-chemical} for details.}
\begin{equation}
\Delta_1=\Delta_2=-\frac{\omega_2}{2}~,
\quad
\Delta_3=-\omega_1~,
\end{equation}
Eq.~\eqref{eq-equiv-int} reduces to
\begin{equation}
    E_C|_{\text{chiral}}= -\frac{1}{4}\bigg(N^2-\sum_{i=1}^{\nu}N_i^2\bigg) \omega_2~,
\end{equation}
consistent with the localization result~\eqref{eq:dsce-schur}.

\bigskip

To conclude, for a GW defect with the moduli parameters $\beta_i = \gamma_i=0$, in 4d $\cN=4$ SYM, we find that the dSCE is governed by the 't~Hooft anomaly $k_\ttR$ of $\SU(2)_\ttR$ symmetry,
\begin{equation}
\label{eq:GW-dSCE-kR}
    E_C=-k_\ttR\frac{\Delta_2 \Delta_3}{2\,\omega_1}~,\quad  k_\ttR=N^2-\sum_{i=1}^{\nu}N_i^2~.
\end{equation}
In this case, the defect Weyl-anomaly coefficients $b$ and $d_2$ are completely determined by the 't~Hooft anomaly $k_\ttR$,
\begin{equation}
    b=d_2=3k_\ttR=3\bigg(N^2-\sum_{i=1}^{\nu}N_i^2\bigg)~.
\end{equation}
Although $b$ and $d_2$ coincide in this case, inspired by the half-BPS surface defect in 6d $\cN=(2,0)$ theory, we expect that the dSCE here is also governed by a linear combination of $b$ and $d_2$. 
\paragraph{dSRE.}
We now calculate the dSRE using the extended integration method.
Before turning to the defect contribution, we first adapt the extended equivariant-integral prescription of~\cite{Yankielowicz:2017xkf}, originally formulated for four-dimensional SCFTs with a $\U(1)$ R-symmetry, to the ambient $\mathcal N=4$ SYM theory with $\SU(4)_R$ R-symmetry. This fixes the extended equivariant parameters and the R-symmetry chemical potentials on the replica background.
For a 4d $\cN=4$ SYM with gauge group $G$, the anomaly polynomial is~\cite{Lawrie:2018jut, Thorngren:2020yht}
\begin{equation}
    I^{(6)} = \frac{d_G}{2}c_3(\mathcal{R})~,
\end{equation}
where $d_G$ is the dimension of $G$, and $c_3(\mathcal{R})$ is the third Chern class of the $\SU(4)_R$ R-symmetry bundle $\mathcal{R}$.
To obtain the SRE from the anomaly polynomial, one extends the ordinary equivariant integral by introducing an auxiliary $\U(1)_T$ rotation with chemical potential $\mu_T$, in addition to the two spacetime rotation parameters $\omega_1,\omega_2$. The supersymmetric constraint is correspondingly extended to
\begin{equation}
\label{eq:ext-BPS}
    \sum_{I=1}^3 \Delta_I+\sum_{i=1}^2\omega_i+\mu_T=0~.
\end{equation}
The extended integral is then taken over
$\mathbb{R}^2_{\omega_1}\times\mathbb{R}^2_{\omega_2}\times\mathbb{R}^2_{\mu_T}$, with $c_3(\mathcal R)$ evaluated equivariantly as\footnote{The equivariant Chern roots of the $\SU(4)_R$ bundle are given by the
weights $x_a$ of a chiral spinor of $\Spin(6)_R\simeq \SU(4)_R$, which are related to $\Delta_I$ as
$
x_1=-\frac12(\Delta_1+\Delta_2+\Delta_3), \,
x_2=\frac12(\Delta_1+\Delta_2-\Delta_3),
x_3 =\frac12(\Delta_1-\Delta_2+\Delta_3), \, 
x_4=\frac12(-\Delta_1+\Delta_2+\Delta_3)
$.
This gives
$c_3(\mathcal{R})=\sum_{a<b<c}x_a x_b x_c=-\Delta_1\Delta_2\Delta_3$.} $c_3(\mathcal{R})\to -\Delta_1 \Delta_2 \Delta_3$.
The extended integral then gives
\begin{equation}
\tilde{\mathsf{E}}^{\mathcal{N} = 4}\qty(\Delta_I,\omega_i,\mu_T) \equiv \int_{\mathbb{R}^2_{\omega_1}\times\mathbb{R}^2_{\omega_2}\times\mathbb{R}^2_{\mu_T} }I^{(6)}=-
\frac{d_G}{2} \frac{\Delta_1 \Delta_2 \Delta_3}{\omega_1\omega_2\mu_T}~.
\end{equation}
The choice of R-symmetry background on the replica geometry depends on the supercharge used to define the SRE. For a 4d $\mathcal N=4$ SCFT, a general background in the Cartan of $\SU(4)_R$ is parametrized as
\begin{equation}
    \Delta_1=-1+s_1\Big(1-\frac{1}{n}\Big)~, \quad \Delta_2=-1+s_2\Big(1-\frac{1}{n}\Big)~, \quad \Delta_3=-1+s_3\Big(1-\frac{1}{n}\Big)~,
\end{equation}
with $s_1 + s_2 + s_3 = 1$. 
Together with $\omega_1=\omega_2=1, \, \mu_T = \frac{1}{n}$, this choice satisfies the extended supersymmetric constraint~\eqref{eq:ext-BPS}.
The SRE is obtained from
\begin{equation}\label{eq:SRE-N4}
\begin{aligned}
\mathsf{S}^{\mathcal{N} = 4}_n &=\frac{n \tilde{\mathsf{E}}^{\mathcal{N} = 4}|_{n=1}-\tilde{\mathsf{E}}^{\mathcal{N} = 4}|_n}{1-n} V_{\mathbb{H}^1}
\\
&=-\frac{d_G}{2}\qty[s_1 s_2 s_3\Big(\frac{1}{n} - 1\Big)^2 + (s_1 s_2 + s_2 s_3 + s_3 s_1)\Big(\frac{1}{n} - 1\Big) + 1]V_{\mathbb{H}^1}~.
\end{aligned}
\end{equation}
This reproduces the SRE of the ambient $\mathcal N=4$ theory computed in~\cite{Huang:2014pda}. 

In the presence of a GW defect, however, the allowed R-symmetry background is restricted by the symmetry preserved by the defect. 
With the GW defect insertion, the unbroken R-symmetry is $\SU(2)_\ttL\times\SU(2)_\ttR$.
Therefore, the allowed background gauge fields are linear combinations of the Cartans $J_\ttL$ and $J_\ttR$, which involve only the bulk R-symmetry Cartans $R_2$ and $R_3$. 
Accordingly, we turn on generic replica-dependent background gauge fields for $R_2$ and $R_3$, while keeping no replica-dependent background along $R_1$.
In the anomaly-polynomial approach to the dSRE, we thus choose
\begin{equation}
\label{eq:chem-2}
\Delta_1 = -1~,\quad \Delta_2 = -1 + s_2 \qty\Big(1-\frac{1}{n})~, \quad \Delta_3= -1 + s_3 \qty\Big(1-\frac{1}{n})~,
\end{equation}
where $s_2 + s_3 = 1$, together with 
\begin{equation}
\label{eq:ext-chem}
\omega_1=\omega_2 =1~, \quad \mu_T=\frac1n~. 
\end{equation}
This choice also satisfies the extended supersymmetric constraint~\eqref{eq:ext-BPS}.
Notice that this choice is compatible with the holographic background~\eqref{eq-holo-cp}.
The extended integral of the defect anomaly polynomial
\begin{equation}
\label{eq:GW-ap}
\I = -\frac{1}{2} \bigg(N^2 - \sum_{i=1}^{\nu} N_i^2\bigg)\qty(c_2(\ttL)-c_2(\ttR))
\end{equation}
then gives
\begin{equation}
\label{eq:GW-edSCE}
\tilde{E} \equiv  \int_{\mathbb{R}^2_{\omega_1}\times\mathbb{R}^2_{\mu_T} }\I= -\frac{k_\ttR}{2}\frac{\Delta_2 \Delta_3}{\omega_1 \mu_T}~,
\end{equation}
Using the setting~\eqref{eq:chem-2} and~\eqref{eq:ext-chem} we obtain
\begin{equation}
\label{eq:GW-dSRE-2}
    S_n=
    \frac{n \tilde{E}|_{n=1}-\tilde{E}|_n}{1-n} V_{\mathbb{H}^1}
=\frac{1}{2}k_\ttR \qty[s_2 s_3\qty\Big(\frac{1}{n} - 1) + 1]V_{\mathbb{H}^1}~.
\end{equation}
Thus, the dSRE is completely determined by the defect 't~Hooft anomaly
$k_\ttR$, while the choice of supersymmetric R-symmetry background enters through $s_2s_3$. 
In particular, for the simple defect labeled by $\bL=\S[\U(1)\times\U(N-1)]$, we have $k_\ttR=2(N-1)$, \eqref{eq:GW-dSRE-2} then gives
\begin{equation}
\label{eq:GW-dSRE-two-equal-charge}
    S_n=(N - 1) \qty[s_2 s_3\qty\Big(\frac{1}{n} - 1) + 1]V_{\mathbb{H}^1}~.
\end{equation}
This result will later be checked by the holographic calculation in Section~\ref{sec:4dGW-holography}. 
\subsection{Holographic dSRE}\label{sec:4dGW-holography}

In this subsection, we compute the dSRE holographically for the simple GW defect with Levi subgroup $\bL=\S[\U(1)\times \U(N-1)]$. The SRE of 4d $\cN=4$ SYM is described holographically by a charged AdS$_5$ topological black hole solution in 5d $\cN=2$ STU gauged supergravity~\cite{Huang:2014pda}. In the presence of the simple GW defect, a natural holographic candidate is a probe D3-brane in the 10d uplift of this black-hole background. The defect expectation value is then obtained from the D3-brane on-shell action, from which we extract the dSRE and find consistency with the anomaly-polynomial result~\eqref{eq:GW-dSRE-two-equal-charge} in Section~\ref{sec:dSRE-int-4d}.

\begin{figure}[htbp]
\centering
\includegraphics[width=1\columnwidth]{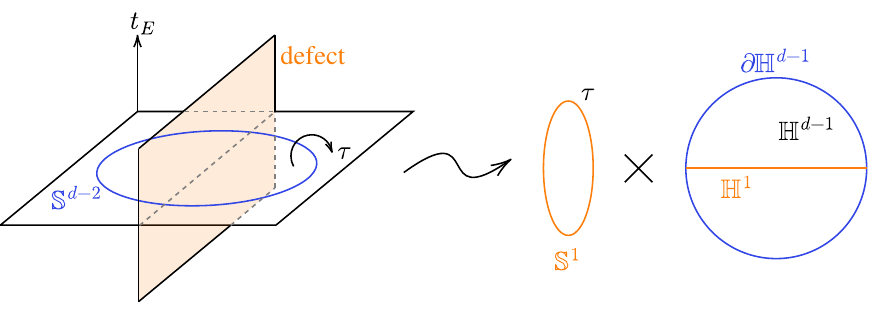}
\caption{A conformal transformation maps the ambient spacetime $\mathbb{R}^d$ to $\mathbb{S}^1 \times \mathbb{H}^{d - 1}$, the planar surface defect $\mathbb{R}^2 \subset \mathbb{R}^d$ to $\mathbb{S}^1 \times \mathbb{H}^1 \subset \mathbb{S}^1 \times \mathbb{H}^{d - 1}$, and the spherical entangling surface $\mathbb{S}^{d - 2}$ to the boundary of $\mathbb{H}^{d - 1}$. After the replica trick, the thermal circle $\mathbb{S}^1$ becomes $\mathbb{S}^1_n$.}
\label{fig-CHM}
\end{figure}

The SRE of the 4d $\cN=4$ SYM is defined on the replica geometry $\bS^1_n \times \bH^3$, with background gauge fields for the R-symmetry Cartan $\U(1)^3\subset \SO(6)_R$ turned on to preserve supersymmetry.
The relevant holographic description is the three-charge hyperbolically sliced AdS$_5$ topological black hole of 5d $\cN=2$ STU gauged supergravity, which is the $\U(1)^3$
Cartan truncation of the 5d maximal $\SO(6)$ gauged supergravity.
The bosonic part of this $\U(1)^3$ truncated gauged supergravity Lagrangian is~\cite{Cvetic:1999xp}\footnote{In this holographic computation, 5d form fields ($A_{(1)}$ and $F_{(2)}$) are denoted by italic letters, with the number in parentheses indicating the degree of the differential form.}
\begin{equation}
\frac{\mathcal{L}_5}{\sqrt{g}} = R - \frac{1}{2}(\partial\vec{\phi})^2 + \frac{4}{L^2}\sum_i X_i^{-1} - \frac{1}{4}\sum_i\frac{\big(F_{(2)}^{i}\big)^2}{X_i^2} + \frac{1}{4}\epsilon^{\mu\nu\rho\sigma\lambda}F_{(2)\mu\nu}^1 F_{(2)\rho\sigma}^2A_{(1)\lambda}^3~,
\end{equation}
where $\vec{\phi} = (\phi_1,\phi_2)$ are two scalars, $F_{(2)}^i = \dd A_{(1)}^i$, and $X_{i = 1,2,3}$ are
\begin{equation}
X_i = e^{-\frac{1}{2}\vec{a}_i\vec{\phi}}~,\quad \vec{a}_1 = \Big(\frac{2}{\sqrt{6}},\sqrt{2}\Big)~,\quad \vec{a}_2 = \Big(\frac{2}{\sqrt{6}},-\sqrt{2}\Big)~, \quad \vec{a}_3 = \Big(-\frac{4}{\sqrt{6}},0\Big)\ .
\end{equation}
This theory admits a three-charge topological black hole solution asymptotic to hyperbolically sliced AdS$_5$ with radius $L$~\cite{Cvetic:1999xp},\footnote{The metric in~\eqref{eq-5d-CTBH} is in Lorentzian signature.}
\begin{align}\label{eq-5d-CTBH}
&\mathrm{d}s_5^2 = -\left(H_1 H_2 H_3\right)^{-2/3} f \mathrm{d}t^2 + \left(H_1 H_2 H_3\right)^{1/3} \left(f^{-1} \mathrm{d}r^2 + r^2 \mathrm{d}s_{\mathbb{H}^3}^2\right)~,\\
&f(r) = -1 - \frac{m}{r^2} + \frac{r^2}{L^2} H_1 H_2 H_3~,\quad H_i = 1 + \frac{q_i}{r^2}~,
\end{align}
together with the scalars and the gauge fields,
\begin{equation}
X_i = \left(H_1 H_2 H_3\right)^{1/3}H_i^{-1}~,\quad A_{(1)}^i = \left(\sqrt{-1+\frac{m}{q_i}} \left( H_i^{-1} - 1 \right) + \hat{\mu}_i \right)\mathrm{d}t~,
\end{equation} 
where $q_i$ are the three charges of the black hole, and $\hat{\mu}_i$ are the associated chemical potentials fixed by the vanishing condition of $A^i_{(1)}$ at the horizon.
In what follows, we restrict to the supersymmetric branch $m=0$. 
The black hole horizon $r_H$, Hawking temperature $T_H$ and the chemical potentials are given by~\cite{Huang:2014pda}
\begin{equation}
r_H = \frac{L}{\sqrt{(1 + \kappa_1)(1 + \kappa_2)(1 + \kappa_3)}}~,
\end{equation}
\begin{equation}
T = \frac{f'(r)}{4\pi \sqrt{H_1 H_2 H_3}}\Big|_{r = r_H} = \frac{1 - \kappa_1 \kappa_2 - \kappa_2 \kappa_3 - \kappa_3 \kappa_1 - 2\kappa_1\kappa_2\kappa_3}{2\pi L(1 + \kappa_1)(1 + \kappa_2)(1 + \kappa_3)}~,
\end{equation}
\begin{equation}\label{eq-holo-cp-k}
 \hat{\mu}_i = A^i_{(1)t}|_{r\to \infty}=\frac{i}{\kappa_i^{-1} + 1}~.
\end{equation}
where we introduce dimensionless charge parameters $\kappa_i = q_i/r_H^2$.
By matching $T_H$ to the boundary temperature $1/(2\pi n)$ on the replica circle $\bS^1_n$, $\hat{\mu}_i$ are identified as the R-symmetry chemical potentials on the boundary, and the holographic SRE is obtained from the black hole Bekenstein-Hawking entropy~\cite{Huang:2014pda}.

The STU black-hole solution admits a standard uplift to ten-dimensional type \Rm{2}B supergravity on $S^5$ \cite{Cvetic:1999xp}.
The uplifted 10d metric is
\begin{equation}
\mathrm{d}s_{10}^2 = \sqrt{\Delta} \mathrm{d}s_5^2 + \frac{1}{g^2 \sqrt{\Delta}} \sum_{i = 1}^3 X_{i}^{-1} \qty(\mathrm{d}\mu_i^2 + \mu_i^2\big(\mathrm{d}\xi_i + g A_{(1)}^i\big)^2)~,
\end{equation}
with $\Delta = \sum_{i = 1}^3 X_i \mu_i^2$, $g = 1/L$, and $\sum_{i = 1}^3 \mu_i^2 = 1$.

Now we consider the insertion of the simple GW defect. On the field theory side, the GW defect wraps $\bS^1_n \times \bH^1 \subset \bS^1_n \times \bH^3$;
see Figure~\ref{fig-CHM} for an illustration of the replica manifold with the surface defect. On the holographic side, a natural candidate is a probe D3-brane in the uplifted type \Rm{2}B supergravity. 
The expectation value of the surface defect is obtained from the on-shell action of the probe D3-brane,
\begin{equation}
I_{\text{D3}} = -\log \langle D\rangle_n \ . 
\end{equation}
The D3-brane action includes both the Dirac-Born-Infeld (DBI) term and the Wess--Zumino (WZ) term,
\begin{equation}
I_{\text{D3}} = T_{3}\int_{M_4} \mathrm{d}^4\sigma \sqrt{-\mathrm{det}[g_{\text{D3}}]} - T_3\int_{M_4} P\big[\mathrm{C}_{4}\big]~,
\end{equation}
where $\sigma$ denotes the worldvolume coordinates of the D3-brane, $g_{\text{D3}}$ is the induced metric on the D3-brane worldvolume $M_4$, and $P[\mathrm{C}_{4}]$ is the pullback of the RR potential $\mathrm{C}_{4}$ on $M_4$. The RR potential $\mathrm{C}_{4}$ satisfies\footnote{In this holographic calculation, 10d form fields ($\mathrm{C}_4$, $\mathrm{F}_5$, $\mathrm{G}_5$) are denoted by upright letters. }
\begin{equation}
\mathrm{d}\mathrm{C}_{4} = \mathrm{F}_{5}~.
\end{equation}
And the 5-form field strength is $\mathrm{F}_{5} = \mathrm{G}_{5} + *\mathrm{G}_{5}$, with $\mathrm{G}_{5}$ given by
\begin{equation}\label{eq-G5}
\begin{split}
\mathrm{G}_{5} =&\, 2g\sum_i (X_i^2 \mu_i^2 - \Delta X_i)\epsilon_{(5)} - \frac{1}{2g}\sum_i X_i^{-1}\bar{*}\mathrm{d} X_i\wedge \mathrm{d}(\mu_i^2)\\
&\,+ \frac{1}{2g^2}\sum_i X_i^{-2} \mathrm{d}\qty(\mu_i^2)\wedge \big(\mathrm{d}\xi_i + g A_{(1)}^i\big)\wedge \bar{*}F_{(2)}^i~,
\end{split}
\end{equation}
where $F_{(2)}^i = \mathrm{d}A_{(1)}^i$, $\epsilon_{(5)}$ is the volume form of $\mathrm{d}s_5^2$ \eqref{eq-5d-CTBH}, and $\bar{*}$ is the Hodge dual with respect to the 5d metric $\mathrm{d}s_5^2$.
In the probe limit, the embedding of D3-brane is 
\begin{equation}\label{eq-D3-embedding}
\sigma^a = (\tau,r,\rho,\xi_1)~.
\end{equation}
In the internal manifold, we place the D3-brane at $(\mu_1, \mu_2, \mu_3) = (1, 0, 0)$, which is compatible with the R-symmetry background~\eqref{eq:chem-2}.\footnote{See~\cite{Crossley:2014oea} for the string embedding case. } It can be checked that when the D3-brane is along $\{\tau,r,\rho,\xi_1\}$, the pullback of $\mathrm{C}_4$ vanishes. Therefore, the on-shell action is\footnote{Here we have Wick rotated to Euclidean signature. }
\begin{equation}
I_{\text{D3}}[n] = T_{3}\int_{\mathbb{S}_n^1} \mathrm{d}\tau \int_{\mathbb{H}^1} \mathrm{d}\rho \int_{r_H}^\Lambda \mathrm{d}r \int_{0}^{2\pi}\mathrm{d}\xi_1 \, r L = 2 N n (\Lambda^2 - r_H^2) \log \frac{\ell}{\epsilon}~,
\end{equation}
where we have used $T_3 = N/(2\pi^2 L^4)$ and set $L = 1$. 
We regulate the asymptotic radial divergence at $r=\Lambda$ by subtracting the term $2Nn\Lambda^2\log(\ell/\epsilon)$. We also note that this term cancels in the combination $nI_{\mathrm{D3}}[1] - I_{\mathrm{D3}}[n]$ entering the dSRE.
We retain the logarithmic factor $\log(\ell/\epsilon)$ and adopt the same radial subtraction prescription in the 6d calculation in Section~\ref{sec:6d-holo}.

To match the chemical potentials of the boundary CFT~\eqref{eq:chem-2}, one should set
\begin{equation}
\label{eq-holo-cp}
\hat{\mu}_1 = 0~, \quad \hat{\mu}_2 = i\qty\Big(1 - \frac{1}{n})s_2~, \quad \hat{\mu}_3 = i\qty\Big(1 - \frac{1}{n})s_3~,
\end{equation}
with $s_2 + s_3 =1$ and here we take $s_1 =0$ since with the defect insertion, the allowed background gauge fields contain the R-symmetry Cartans $R_2$ and $R_3$ only. Combining \eqref{eq-holo-cp} and \eqref{eq-holo-cp-k} gives
\begin{equation}
r_H^2 = \qty(\!1-\qty\Big(1-\frac{1}{n})s_2\!)\qty(\!1-\qty\Big(1-\frac{1}{n})s_3\!)~,\quad I_{\text{D3}}[n] = -2 N \frac{s_2 s_3(n - 1)^2 + n}{n}\log \frac{\ell}{\epsilon}~.
\end{equation} 
As a consistency check, at $n=1$ we obtain 
\begin{equation}
\log\langle D\rangle = - I_{\text{D3}} = 2N\log\frac{\ell}{\epsilon}~,
\end{equation}
which agrees with the expected universal contribution $(b/3)\log \ell/\epsilon$~\cite{Kobayashi:2018lil, Jensen:2018rxu} for the simple defect labeled by $\bL=\S[\U(1)\times\U(N-1)]$, whose anomaly coefficient is $b=6(N-1)\sim 6N$ at large $N$. The dSRE is
\begin{equation}
\label{eq:GW-dSRE-holography}
S_n = \frac{\log \langle D\rangle_n - n \log\langle D\rangle}{1 - n} = 2N\qty[s_2 s_3\qty\Big(\frac{1}{n} - 1) + 1]\log\frac{\ell}{\epsilon}~,
\end{equation}
which is consistent with the anomaly polynomial result~\eqref{eq:GW-dSRE-two-equal-charge} at large $N$ limit.
\section{\boldmath Half-BPS Surface defects in 6d SCFTs\unboldmath}\label{sec:6d}
In this section we turn to half-BPS surface defects in 6d superconformal theories. 
Six-dimensional superconformal field theories provide an important class of strongly coupled quantum field theories, which in general do not admit Lagrangian descriptions~\cite{Heckman:2018jxk}. Nevertheless, broad classes of 6d SCFTs can be constructed and classified geometrically in F-theory~\cite{Heckman:2013pva,Heckman:2015bfa}, and can also be engineered through brane configurations in M-theory~\cite{DelZotto:2014hpa} and type IIA massive string theory~\cite{Hanany:1997gh}. These constructions provide useful descriptions and tools to explore the 6d SCFTs and the extended objects.

Surface defects provide useful probes of the non-perturbative structure of 6d SCFTs.
They admit complementary descriptions in terms of brane constructions [8, 31] and coupled 2d–6d systems.
In 6d $\mathcal N=(1,0)$ SCFTs, a rich class of half-BPS surface defects preserve two-dimensional $(0,4)$ supersymmetry. Their chiral degrees of freedom can give rise to a nonzero defect gravitational anomaly.

In this section, we derive general expressions for the dSCE and dSRE from the defect anomaly polynomial and relate them to the defect anomaly coefficients. We then apply these results to two representative classes of defects: the surface defects engineered by D4-branes in type \Rm{2}A constructed in~\cite{Apruzzi:2024ark} in Section~\ref{sec:large-04-6d}, and half-BPS Wilson surfaces in 6d $\cN=(2,0)$ SCFTs
in Section~3.3.
\subsection{General results}\label{sec:general-results}
In this section we consider a half-BPS superconformal surface defect in 6d SCFTs with at least $\cN=(1,0)$ supersymmetry. We first list the symmetry structure in the presence of the defect, and then use the anomaly polynomial method to obtain the general results of dSCE and dSRE.
\paragraph{The symmetry structure.}
The half-BPS surface defect
breaks the six-dimensional conformal symmetry as 
\[
\SO(6,2)\,\longrightarrow\, \SO(2,2)\times \SO(4)_{\text{perp}}~,
\]
where $\SO(2,2)$ is the conformal symmetry along the defect, and
\[
\SO(4)_{\text{perp}}\simeq \SU(2)_L\times \SU(2)_R
\]
is the rotation symmetry in the transverse directions.
The defect also preserves the $\SU(2)_I$ R-symmetry of the ambient 6d $\cN=(1,0)$ theory.

The surface defect can preserve \textit{large} or \textit{small} 2d $(0,4)$ supersymmetry.
For large $(0,4)$ supersymmetry, the defect R-symmetry is $\SU(2)_I\times\SU(2)_R$, while $\SU(2)_L$ acts as a flavor symmetry.
For small $(0,4)$ supersymmetry, only $\SU(2)_I$ is the R-symmetry, while both $\SU(2)_L$ and $\SU(2)_R$ act as flavor symmetries.

To make the embedding of the defect superconformal symmetry into the six-dimensional one explicit, we now specify the corresponding Cartan generators.
We denote by $E$ the dilatation generator, $J_1,J_2,J_3$ the three Cartans of the $\SO(6)$ rotation symmetry, and $J_I$ the Cartan of the $\SU(2)_I$ R-symmetry.
Without loss of generality, we choose the defect to lie in the two-plane whose rotation is generated by $J_1$, while $J_2$ and $J_3$ generate rotations in the two transverse planes.
The Cartan generators of the 2d conformal symmetry and the preserved $\SU(2)_L\times\SU(2)_R$ symmetry are then identified as
\begin{equation}\label{eq:Cartan-idf}
    L_0=\frac{E+J_1}{2}~, \quad
    \bar{L}_0=\frac{E-J_1}{2}~, \quad
    J_L=\frac{J_2+J_3}{2}~, \quad
    J_R=\frac{J_2-J_3}{2}~.
\end{equation}
Here $L_0$ and $\bar L_0$ are the left- and right-moving conformal Hamiltonians on the defect, while $J_L$ and $J_R$ are the Cartan generators of $\SU(2)_L$ and $\SU(2)_R$, respectively.
\paragraph{The anomaly polynomial method.}
Having established the preserved symmetries and their embedding into the ambient theory, we now use the defect anomaly polynomial to determine the dSCE and dSRE. The preserved symmetry structure fixes the general form of the defect anomaly polynomial,
\begin{equation}
\label{eq:6d-defect-ap-general}
    \I = \frac{1}{2}k_L c_2(L) + \frac{1}{2}k_R c_2(R) + \frac{1}{2}k_I c_2(I) - \frac{k_g}{24} p_1(T)~,
\end{equation}
where $k_{L,R,I}$ are the defect 't~Hooft anomalies of the preserved $\SU(2)_{L,R,I}$ symmetry, $k_g$ is the defect gravitational anomaly.\footnote{Possible additional global symmetries are omitted for simplicity. If present, they can be included by adding the corresponding characteristic classes to the defect anomaly polynomial.}
The equivariant integral of the defect anomaly polynomial~\eqref{eq:6d-defect-ap-general} gives
\begin{equation}
\label{eq:dSCE-6d(1,0)-general}
\int_{\mathbb{R}^2_{\omega_\ell}} \!\!\I = -\frac{1}{8\omega_\ell}\qty(k_L a_L^2 + k_R a_R^2 + k_I a_I^2 ) - \frac{1}{24}k_g \omega_\ell~,
\end{equation}
where $\omega_\ell$ denotes the equivariant parameter of the 2d plane occupied by the surface defect. 

Consider the pullback corresponding to the defect-preserved global symmetry,
\begin{equation}
\label{eq:pullback}
    a_I J_I + a_R J_R+a_L J_L=\sigma J_I+\omega_2 J_2+\omega_3 J_3~.
\end{equation}
Using the dictionary of Cartan generators~\eqref{eq:Cartan-idf}, the chemical potentials are identified as
\begin{equation}
\label{eq:chem-pot-dict}
 \omega_\ell=\omega_1~, \quad  a_I = \sigma~,  \quad a_L = \omega_2 + \omega_3~,\quad 
    a_R = \omega_2 - \omega_3~.
\end{equation}
The equivariant integral of the defect anomaly polynomial~\eqref{eq:6d-defect-ap-general} then gives the dSCE,
\begin{equation}\label{eq:dSCE-6d(1,0)}
E_C 
= -\frac{1}{8\omega_1}\qty(k_L(\omega_2 + \omega_3)^2 + k_R(\omega_2 - \omega_3)^2 + k_I(\omega_1 + \omega_2 + \omega_3)^2 ) - \frac{1}{24}k_g \omega_1~,
\end{equation}
where we have used the BPS condition $\sigma=\sum_{i=1}^3 \omega_i$. 

\bigskip

We now turn to the anomaly polynomial approach to dSRE. 
Since $\SU(2)_I$ remains an R-symmetry of the defect, the same replica-dependent $\SU(2)_I$ background is allowed in the presence of the defect. We therefore extend the defect anomaly polynomial by introducing an extended $\U(1)$ with chemical potential $\mu_T$, as well as the replacement rule 
\begin{equation}
\label{eq:6d-replacement}
    \sigma \to \sum_{i=1}^3 \omega_i + \mu_T~, \quad p_1(T) \to \omega_1^2+\mu_T^2~,
\end{equation}
together with the setting 
\begin{equation}
\label{eq:6d-extended-setting}
    \omega_1=\omega_2=\omega_3=1~, \quad \mu_T=\frac{1}{n}~.
\end{equation}
It is clear that the chemical potential $a_I=\sigma$ has $1/n$-dependence, implying that the background gauge field of $\SU(2)_I$ R-symmetry is turned on to preserve supersymmetry.
The extended integral of the defect anomaly polynomial gives
\begin{equation}
\begin{aligned}
\tilde{E} \equiv & \, \int_{\mathbb{R}^2_{\omega_1}\times\mathbb{R}^2_{\mu_T} }\I \\
= &\, -\frac{1}{8\omega_1 \mu_T}k_I(\omega_1 + \omega_2 + \omega_3 + \mu_T)^2 - \frac{k_g}{24\omega_1 \mu_T}\qty(\omega_1^2 + \mu_T^2)\\
&\,-\frac{1}{8\omega_1 \mu_T}\qty( k_L(\omega_2+\omega_3)^2+k_R(\omega_2-\omega_3)^2)~,
\end{aligned}
\end{equation}
from which we obtain the dSRE,
\begin{equation}
\label{eq:dSRE-6d(1,0)-general}
S_n = \frac{n\tilde{E}|_{n = 1} - \tilde{E}|_n}{1 - n}V_{\mathbb{H}^1} = \frac{1}{24}\qty(\frac{3k_I + k_g}{n} + 21k_I + k_g)V_{\mathbb{H}^1}~.
\end{equation}
It exhibits the dependence on the gravitational anomaly,\footnote{In theories with a gravitational anomaly, the choice of cutoff prescription could introduce an ambiguous finite term in the entanglement entropy~\cite{Iqbal:2015vka}.
Here we focus only on the universal logarithmic term.}
\begin{equation}
    S_n \supset \frac{k_g}{24}\qty(\frac{1}{n}+1)V_{\mathbb{H}^1}~.
\end{equation}
For the $\cN=(1,0)$ supersymmetric background considered here, the dSRE contains a universal contribution controlled by the $\SU(2)_I$ anomaly,
\begin{equation}\label{eq-kI-dSRE}
S_n \supset \frac{k_I}{8}\qty(\frac{1}{n}+7)V_{\mathbb{H}^1}~.
\end{equation}
This is because $\SU(2)_I$ is the R-symmetry group turned on to preserve the supersymmetry.
Especially, for $\cN=(2,0)$ theories, the R-symmetry is enhanced to $\SO(5)_R$, and the identification of the R-symmetry chemical potentials requires a slight modification. In terms of $\SU(2)_I \times \SU(2)_F \subset \SO(5)_R$, a generic supersymmetric R\'enyi background may involve both Cartans. The $\cN=(1,0)$ background used above corresponds to the specialization in which only the $\SU(2)_I$ Cartan carries replica dependence. We return to this point in Section~\ref{sec:6d-BPS-string}.
\subsection[Surface defects in 6d $(1,0)$ SCFTs]{\boldmath Surface defects in 6d $(1,0)$ SCFTs\unboldmath}\label{sec:large-04-6d}
In this section, we consider a concrete example of a half-BPS surface defect in 6d $\cN=(1,0)$ SCFTs constructed by D4-D2 branes in type \Rm{2}A~\cite{Apruzzi:2024ark}. For this defect, we first specialize the general anomaly-polynomial results of Section~\ref{sec:general-results} to obtain the dSCE and dSRE in terms of the defect R-symmetry anomaly $k_I$ and gravitational anomaly $k_g$. We then compute dSCE and dSRE independently using the following complementary descriptions: supersymmetric localization in the 5d circular-quiver theory, the 2d localized chiral sector of the defect, and a holographic probe D4-brane calculation.

The full bulk-defect system is engineered by the near-horizon limit of an NS5-D6-D8 brane system in the 10d Type \Rm{2}A supergravity. Consider $N$ NS5-branes located in the $x^{0,\cdots,5}$ directions, with $r_i$ D6-branes suspended between the $(i-1)$-th and the $i$-th NS5-branes. The D6-branes occupy the $x^{0,\cdots,6}$ directions. There are also $f_i=2r_i-r_{i+1}-r_{i-1}$ D8-branes extending across the $r_i$ D6-branes, occupying the $x^{0,\cdots,5,7,8,9}$ directions. The surface defects are engineered by additional $q_i$ D4-branes along the $x^{0,1,7,8,9}$ directions perpendicular to the $r_i$ D6-branes.
The defect can also be described by coupling a 2d quiver theory to the 6d one, with the 2d gauge nodes and multiplets determined by the brane structure.
For the detailed construction, see~\cite{Dibitetto:2017klx,Apruzzi:2024ark}.

The defect preserves large 2d $\cN=(0,4)$ supersymmetry. In the notation of Section~\ref{sec:general-results}, its R-symmetry is $\SU(2)_I\times\SU(2)_R$, while $\SU(2)_L$ acts as a flavor symmetry.
The defect anomaly polynomial is computed in~\cite{Apruzzi:2024ark},
\begin{equation}
\label{eq:defect-ap-(1,0)}
    I^{(4)}
    =
     \frac{k_I}{2} c_2(I)-\frac{k_g}{24}p_1(T)~,
\end{equation}
with the defect anomaly coefficients\footnote{Notice that our sign convention is opposite to~\cite{Apruzzi:2024ark}. Here we define the gravitational anomaly as $k_g=c_L-c_R$, while their convention is $k_g^{(\text{their})}=c_R-c_L$.} 
\begin{equation}
k_g = \sum_i q_i r_i~, \quad k_I= 2q_i(C^{-1})^{ij} r_j~,
\end{equation}
and $C_{ij}$ is the Cartan matrix of $A_{N-1}$.
Using the equivariant-integration prescription of Section~\ref{sec:general-results}, the defect anomaly polynomial~\eqref{eq:defect-ap-(1,0)} leads to the dSCE,
\begin{equation}
\label{eq:dSCE-ap-6d(1,0)}
E_C=\int_{\mathbb{R}^2_{\omega_1}} \!\!\I = -\frac{1}{8\omega_1} k_I (\omega_1+\omega_2+\omega_3)^2-\frac{1}{24} k_g \omega_1~,
\end{equation}
universally controlled by the defect 't~Hooft anomaly $k_I$ and gravitational anomaly $k_g$.
Similarly, the dSRE is obtained from the extended integral of the defect anomaly polynomial~\eqref{eq:defect-ap-(1,0)} by following the replacement rules~\eqref{eq:6d-replacement} and~\eqref{eq:6d-extended-setting}, 
\begin{equation}
\label{eq:defect-6d(1,0)-dSRE}
S_n = \frac{1}{24}\qty(\frac{3k_I + k_g}{n} + 21 k_I + k_g)V_{\mathbb{H}^1}~. 
\end{equation}

In what follows, we provide consistency checks of this formula.
In Section~\ref{sec:6d-10-localization}, we compute the $k_I$-dependent piece in dSCE using supersymmetric localization and find complete agreement with the anomaly-polynomial prediction. In Section~\ref{sec:6d-10-2d-quiver}, we use the 2d quiver description of the defect to determine the $k_g$-dependent contributions to both the dSCE and the dSRE, again reproducing the general results. In Section~\ref{sec:6d-holo}, we perform a holographic probe-brane computation and recover the $k_I$-dependent piece of the dSRE.
\subsubsection[Wilson loop in the 5d $\cN=1$ circular quiver]{\boldmath Wilson loop in the 5d $\cN=1$ circular quiver\unboldmath}\label{sec:6d-10-localization}
In this subsection, we extract the $k_I$-dependent contribution to the dSCE using Wilson loops in the 5d $\cN=1$ circular quiver theory obtained by compactifying the 6d SCFT on $\mathbb{S}^1_\beta$.
We focus on the massless type IIA case, $F_0=0$, with $r_i=k$ for all $i$. In this limit, the type IIA background uplifts to M-theory on the $\mathbb{C}^2/\mathbb{Z}_k$ orbifold, and the $\mathbb{S}^1_\beta$ reduction of the 6d theory gives a 5d $\cN=1$ circular quiver with $k$ $\SU(N)$ gauge nodes, with a massless bifundamental hypermultiplet between each pair of adjacent nodes.

Upon reduction along $\mathbb{S}^1_\beta$, the surface defect wrapping $\mathbb{S}^1_\beta\times \mathbb{S}^1_{\omega_\ell}$ reduces to a line defect wrapping $\mathbb{S}^1_{\omega_\ell}\subset \mathbb{S}^5_{\vec{\omega}}$ in the 5d theory.  
For a single surface defect insertion, $q_j=1$ and $q_{i\neq j}=0$, we consider the corresponding line operator as the cyclically invariant average of rank-$j$ antisymmetric Wilson loops in this quiver theory. 
A related Wilson-loop computation on the round five-sphere was presented in~\cite{Apruzzi:2024ark}.
Here we instead work directly with the matrix model of the 5d $\cN=1$ $\SU(N)^k$ circular quiver at generic squashing parameters $\omega_{1,2,3}$, with the gauge couplings related to the six-dimensional compactification circumference $\beta$ through~\eqref{eq:quiver-coupling-dictionary}.

We first review the $\mathbb{S}^5$ partition function of the 5d $\cN=1$ theory with gauge group $G$. 
Consider the large rank limit so that the instanton contribution can be suppressed. The perturbative partition function on the squashed five-sphere $\mathbb{S}^5_{\vec\omega}$ takes the form~\cite{Kallen:2012va, Imamura:2012efi, Minahan:2016xwk}
\begin{equation}
\label{eq:5d-matrix-model}
    Z_{\text{pert}}
    =
    \int
    \prod_{i=1}^{\text{rank}\, G} \dd a_i
    \,Z_{\rm cl}(a)
    Z_{\rm 1-loop}(a)~,
\end{equation}
where $a$ is the scalar vev in the Cartan of the gauge group, and
\begin{equation}
    Z_{\rm cl}(a)=\exp\qty(-\frac{4\pi^3}{g_{\text{YM}}^2 \omega_1\omega_2\omega_3}\Tr a^2)
\end{equation}
is the classical contribution, and
\begin{equation}
    Z_{\rm 1-loop}(a)= \frac{\prod_{\sigma \in \text{root}}S_3(-i \sigma(a),\vec\omega)}{\prod_{\rho \in \mathcal{R}}S_3\qty(-i \rho(a)+\frac{\omega_1+\omega_2+\omega_3}{2},\vec\omega)}
\end{equation}
is the 1-loop contribution from the vector multiplet and the massless hypermultiplet. $S_3(z, \vec\omega)$ denotes the triple sine function. 
$\sigma$ runs over all the roots of the gauge group $G$, $\mathcal{R}$ is the representation of the hypermultiplet in the gauge group $G$, and $\rho$ is the weight of the representation $\mathcal{R}$. 

We now generalize the partition function of the 5d $\cN=1$ theory with a single gauge group to a circular quiver theory with $k$ gauge nodes, the gauge groups being $G^{(\alpha)}=\SU(N), \ \alpha=1,\dots,k$. The field content consists of one vector multiplet $V^{(\alpha)}$ for each gauge node, and one bi-fundamental massless hypermultiplet $\Phi^{(\alpha,\alpha+1)}$ for each two adjacent gauge groups, transforming in the representation $(\bold{N}_\alpha,\bold{\bar N}_{\alpha+1})$, $\alpha=1,\dots,k$.
The partition function is given by a generalization to $k$ nodes of~\eqref{eq:5d-matrix-model},
\begin{equation}
\label{eq:5d-n1-quiver}
    Z_{\mathbb{S}^5_{\vec\omega}}
    =
    \int
    \prod_{\alpha=1}^{k} \qty{ \qty[
    \prod_{i=1}^{N}\dd a_i^{(\alpha)}
    \,
    \exp\!\bigg(\!-\frac{4\pi^3}{\big(g_{\text{YM}}^{(\alpha)}\big)^2 \omega_1\omega_2\omega_3}    \Big(a_i^{(\alpha)}\Big)^2\bigg)
    ]
    Z_{\rm 1-loop}^{(\alpha)}}~,
\end{equation}
where $a_i^{(\alpha)}$ are the scalar vevs for each gauge node, satisfying the tracelessness condition $\sum_{i=1}^N a_i^{(\alpha)}=0$. The 1-loop determinant at the $\alpha$-th node receives contribution from the vector multiplet $V^{(\alpha)}$ and the bifundamental hypermultiplet $\Phi^{(\alpha,\alpha+1)}$,
\begin{equation}
    Z_{\rm 1-loop}^{(\alpha)}
    =
    Z_{\rm vec}^{(\alpha)}
    Z_{\rm bif}^{(\alpha,\alpha+1)}~,
    \label{eq:one-loop-total}
\end{equation}
where the vector multiplet contributes
\begin{equation} \label{eq:vector-det}
    Z_{\rm vec}^{(\alpha)}
    =
    \prod_{i\neq j}
    S_3\qty(
        -i\qty(a_i^{(\alpha)}-a_j^{(\alpha)})
        , \vec\omega
    )~,
\end{equation}
and the bifundamental hypermultiplet between the $\alpha$-th and $(\alpha+1)$-th node contributes
\begin{equation}
    Z_{\rm bif}^{(\alpha,\alpha+1)}
    =
    \prod_{i,j=1}^{N}
    S_3\!\left(
        -i\left(a_i^{(\alpha)}-a_j^{(\alpha+1)}\right)+\frac{\omega_1+\omega_2+\omega_3}{2}, \vec\omega
    \right)^{-1}~.
    \label{eq:bif-det}
\end{equation}
In general, the gauge couplings $g_{\text{YM}}^{(\alpha)}$ can differ from node to node.
For the circle reduction considered here, the circumference $\beta$ fixes the sum of their inverses through~\cite{Haghighat:2013tka},
\begin{equation}
    \sum_{\alpha=1}^{k}
    \frac{1}{\big(g_{\text{YM}}^{(\alpha)}\big)^2}
    =
    \frac{1}{2\pi\beta}~.
    \label{eq:quiver-coupling-dictionary}
\end{equation}
As we are working on a particular 5d quiver theory with M-theory construction on $\mathbb{C}^2/\mathbb{Z}_k$, the $\mathbb{Z}_k$ orbifold symmetry naturally motivates choosing equal gauge couplings at all nodes:
\begin{equation}
    \big(g_{\text{YM}}^{(\alpha)}\big)^2
    =
    2\pi k\beta~,
    \quad \alpha=1,\dots,k~.
    \label{eq:equal-node-couplings}
\end{equation}
We now evaluate the integral~\eqref{eq:5d-n1-quiver} in the strong coupling limit, $\beta\to \infty$ in this symmetric gauge node setting. We restrict to a large-$N$ saddle invariant under cyclic permutations of the nodes, $a_i^{(\alpha)}\equiv a_i$.
We also adopt the working assumption that in the $\beta \to \infty$ limit the eigenvalues $a_i$ are very large, $a_i \sim \mathcal{O}(\beta)$, while the chemical potentials remain of order one.
Using the asymptotic behaviours of the triple sine function, (see~\cite[Eq.~(99)]{Imamura:2012efi} and~\cite[Eq.~(3.26)]{Bobev:2015kza}),
\begin{equation}
\begin{aligned}
    \log S_3\!\left(
        ix | \vec\omega
    \right)
    =&-\frac{\pi}{6\omega_1\omega_2\omega_3}|x|^3 + \frac{i\pi(\omega_1+\omega_2+\omega_3)}{4\omega_1\omega_2\omega_3}x|x| \\
    &+ \frac{\pi 
    (
    \omega_1^2 + \omega_2^2 + \omega_3^2 + 
    3
    \qty( 
    \omega_1\omega_2 + \omega_2\omega_3 + \omega_3\omega_1
    )
    )}{12\omega_1\omega_2\omega_3}|x| + \mathcal{O}(1)~,\\
\end{aligned}
\end{equation}
\begin{equation}
    \log S_3\!\left(
       ix+ \frac{\omega_1+\omega_2+\omega_3}{2}
        \Big| \vec\omega
    \right)
    =-\frac{\pi}{6\omega_1\omega_2\omega_3}|x|^3 - \frac{\pi (\omega_1^2 + \omega_2^2 + \omega_3^2)}{24\omega_1\omega_2\omega_3}|x| + \mathcal{O}(1)~,
\end{equation}
the leading effective action for each gauge node evaluated on the cyclically symmetric configuration is
\begin{equation}
    S_{\rm eff}
    =
        \frac{2\pi^2}{\beta \omega_1\omega_2\omega_3}
        \sum_{i=1}^{N} a_i^2
        -
        k\frac{\pi(\omega_1+\omega_2+\omega_3)^2}{8\omega_1\omega_2\omega_3}
        \sum_{i\neq j}|a_i-a_j|~.
    \label{eq:effective-action-general-squashing}
\end{equation}
Ordering the eigenvalues as $a_1<a_2<\cdots<a_N$, the discrete saddle reads
\begin{equation}\label{eq:discrete-saddle}
    a_i^\star
    =
\frac{\beta k(\omega_1+\omega_2+\omega_3)^2}{16\pi}(2i-N-1)~.
\end{equation}
Evaluating the effective action on the saddle gives
\begin{equation}
 \label{eq:free-energy-general-squashing}
    F_{\vec\omega}
    \equiv
    -\log Z_{\mathbb{S}^5_{\vec\omega}}
    \sim
    -\frac{\beta k^2(\omega_1+\omega_2+\omega_3)^4}{384\omega_1\omega_2\omega_3}
    N(N^2-1)~,
\end{equation}
which agrees with the leading contribution obtained by applying the anomaly-polynomial method~\cite{Bobev:2015kza} to the anomaly polynomial derived in~\cite{Ohmori:2014kda}. 

We now consider a half-BPS Wilson loop wrapping the great circle $\mathbb{S}^1_{\omega_\ell} \subset\mathbb{S}^5_{\vec\omega}$.
At the $\alpha$-th gauge node, the Wilson loop in the rank-$j$ antisymmetric representation $[j]$ is
\begin{equation}
    \label{eq:antisymmetric-wilson-character}
    W_j^{(\alpha,\ell)}
    =
    \Tr_{[j]}e^{2\pi a^{(\alpha)}/\omega_\ell}
    =
    \sum_{1\leq i_1<\cdots<i_j\leq N}
    \exp\!\left[
        \frac{2\pi}{\omega_\ell}
        \sum_{s=1}^{j}a_{i_s}^{(\alpha)}
    \right]~.
\end{equation}
The cyclic symmetry of the $k$-node quiver implies
\begin{equation}
    \left\langle W_j^{(\alpha,\ell)}\right\rangle = \left\langle W_{j}^{(\ell)}\right\rangle~,
    \quad \alpha=1,\dots,k~.
    \label{eq:wilson-node-equality}
\end{equation}
It is therefore natural to consider the $\mathbb{Z}_k$-invariant Wilson loop by averaging over the gauge nodes\footnote{Such node-averaged Wilson loops naturally arise in circular quiver theories~\cite{Rey:2010ry}. },
\begin{equation}
   W_{j,\mathrm{cyc}}^{(\ell)}
    \equiv
    \frac{1}{k}\sum_{\alpha=1}^{k}
    W_j^{(\alpha,\ell)}~.
    \label{eq:wilson-untwisted-definition}
\end{equation}
In the probe limit, with $j$ fixed as $N\to\infty$, the Wilson loop insertion does not modify the leading saddle-point eigenvalue distribution~\eqref{eq:discrete-saddle}.
Since the $a_i^\star=\mathcal{O}(\beta)$, the leading large-$\beta$ contribution comes from the largest $j$ eigenvalues,
\begin{equation}
    W_{j}^{(\ell)}
    \simeq \exp \!\left[
    \frac{2\pi}{\omega_\ell}
    \sum_{i=N-j+1}^{N}a_i^\star
    \right]~.
    \label{eq:wilson-leading}
\end{equation}
Without loss of generality, we assume that the Wilson loop is inserted in the great circle labelled by $\ell=1$. The free energy of the cyclically invariant Wilson loop is then evaluated as 
\begin{equation}
    \label{eq:wilson-untwisted}
    \log
    \expval{W_{j,\mathrm{cyc}}^{(\ell=1)}}
    \sim
    \frac{\beta(\omega_1+\omega_2+\omega_3)^2}{8\omega_1}
    kj(N-j)~.
\end{equation}
The large $N$ leading piece of the dSCE is then extracted as  
\begin{equation}
\label{eq:dSCE-(1,0)-lit}
    E_C=-\frac{(\omega_1+\omega_2+\omega_3)^2}{8\omega_1}
        kj(N-j) = -\frac{(\omega_1+\omega_2+\omega_3)^2}{8\omega_1}k_I~,
\end{equation}
with $k_I=kj(N-j)$ in this case.\footnote{$k_I=2q_i(C^{-1})^{il}r_l=2k\sum_i(C^{-1})^{ij}= kj(N-j)$.}
This precisely reproduces the $k_I$-dependent term in the
anomaly-polynomial result~\eqref{eq:dSCE-ap-6d(1,0)}.
The $k_g$-dependent piece, $-k_g\omega_1/24$, is subleading in the large-$N$ expansion and is therefore not captured by the leading saddle-point evaluation of the localized Wilson-loop insertion $\Tr_R e^{2\pi a/\omega_\ell}$. Recovering the $k_g$-dependent term from localization would require keeping additional contributions to the defect partition function.
\subsubsection{The 2d localized chiral sector}\label{sec:6d-10-2d-quiver}
In what follows, we use the 2d quiver theory description of the surface defect to extract the defect gravitational anomaly $k_g$-dependence in the dSCE and dSRE. 

On the tensor branch, the holographic NS5-D6-D8 brane construction gives a six-dimensional linear quiver gauge theory with gauge groups $\U(r_i)$.
The additional defect D4-D2 branes give rise to a 2d $\cN=(0,4)$ quiver coupled to the 6d theory through gauging the
$\U(r_i)$ symmetries~\cite{Apruzzi:2024ark}.
Here we omit the complete field content of the 2d quiver theory since it is complicated; readers can refer to~\cite[Appendix B]{Apruzzi:2024ark} for details.

The full 2d quiver construction includes both the surface defect associated with the D4-brane sector and an additional D2-brane sector in the 10d Type \Rm{2}A setup.
Since we consider only the surface
defect here, we retain the multiplets associated with the D4-brane
degrees of freedom. The D4-D6 open strings give rise to 2d localized $(0,4)$ Fermi multiplets in the bi-fundamental representation $(\bold{q_i},\bold{r_i})$ of $\U(q_i)\times\U(r_i)$, $i=1,\dots,N$. These 2d localized fermions are coupled to the 6d ambient theory by gauging $\U(r_i)$, while there is no self-interaction among these 2d fermions. 
At fixed 6d gauge background $a$, integrating out the 2d localized fermions $\psi$ gives their contribution to the partition function as a fermionic determinant $Z_{\text{det}}$. 

For the supersymmetric partition function, the defect is inserted on $\bS^1_\beta\times \bS^1_{\omega_\ell}\subset \bS^1_\beta\times \bS^5_{\vec\omega}$. The fermions have the anti-periodic boundary condition (NS-sector) along the spatial circle $\bS^1_{\omega_\ell}$, and the boundary conditions along the time circle $\bS^1_\beta$ are twisted by the 2d R-symmetry~\cite{Bobev:2015kza}. As the 2d localized fermionic fields are neutral under the 2d R-symmetry~\cite{Apruzzi:2024ark}, they have the periodic boundary condition (R-sector) along the Euclidean time circle. Therefore, the 2d determinant $Z_{\text{det}}$ is given by
\begin{equation}
\label{eq:EG-2d-quiver}
    Z(\tau,u,v)=\prod_{i=1}^N \prod_{a=1}^{r_i}\prod_{b=1}^{q_i} \frac{\theta_4\qty\big(\tau,u_a^{(i)}-v_b^{(i)})}{\eta(\tau)}~,
\end{equation}
where $u_a^{(i)}, v_b^{(i)}$ are the holonomies of $\U(r_i), \U(q_i)$, respectively, and we use the NS-R sector contribution of a complex left-moving fermion as
\begin{equation}
    Z_{\text{Fermi}}(\tau,z)=\prod_{\rho\in\text{weights}}\frac{\theta_4\qty(\tau,\rho(z))}{\eta(\tau)}~,
\end{equation}
with $\eta(\tau)$ the Dedekind eta function, $\theta_4(\tau,z)$ the Jacobi theta function,
\begin{equation}
\eta(\tau)=q^{1/24}\prod_{n=1}^\infty(1-q^n)~, \quad \theta_4(\tau,z)=\sum_{n\in \mathbb{Z}} (-1)^n q^{\frac{1}{2}n^2}e^{2\pi i n z}~,
\end{equation}
and $q=e^{2\pi i \tau} =e^{-\beta\omega_\ell}$.
In the low-temperature limit, $\beta \to\infty$, $u,v$ fixed, one can extract the leading behaviour of~\eqref{eq:EG-2d-quiver},
\begin{equation}
\label{eq:dSCE-2d-quiver}
    Z(\tau,u,v)\sim q^{-\frac{1}{24}\sum_i q_i r_i}=e^{-\beta \qty(-\frac{1}{24}\sum_i q_i r_i)\omega_\ell}~,
\end{equation}
which recovers the $k_g$-dependent piece of dSCE, $-\frac{1}{24}k_g\omega_\ell$, in the anomaly polynomial result~\eqref{eq:dSCE-ap-6d(1,0)}.

The gravitational piece of the dSRE can also be obtained from the 2d localized sector.
In the context of the dSRE, one needs to compute the thermal partition function on $\mathbb{S}^1_n\times \mathbb{H}^5$, with the defect inserted on $\mathbb{S}^1_n\times \mathbb{H}^1 \subset \mathbb{S}^1_n\times \mathbb{H}^5$. 
As $\mathbb{H}^1$ can be approximated by an infinitely large circle, the 2d fermionic determinant can be computed by that on $\mathbb{S}^1_L \times \mathbb{S}^1_n$ with $L\to \infty$. The 2d fermions have anti-periodic boundary conditions (NS-NS sector) along both $\mathbb{S}^1_L$ and the replica circle $\mathbb{S}^1_n$.\footnote{In principle, the fermionic boundary condition along $\mathbb{S}^1_n$ is twisted by the R-symmetry charge of the fermions. However, the 2d chiral fermions considered here are neutral under the R-symmetry and hence the anti-periodic boundary condition along the replica circle $\mathbb{S}^1_n$ is not twisted.} We also turn off the holonomies $u,v$. Thus the 2d determinant is
\begin{equation}\label{eq:Zn}
    Z_n\equiv Z(\tau_n)= \qty(
\frac{\theta_3\qty(\tau_n)}{\eta(\tau_n)})^{\sum_{i=1}^N q_i r_i} = \qty(
\frac{\theta_3\qty(\tau_n)}{\eta(\tau_n)})^{k_g}~,
\end{equation}
where $\tau_n=iL/2\pi n$, and $\theta_3(\tau,z)=\sum_{m\in \mathbb{Z}} e^{\pi i\tau m^2+2\pi i m z}$. Expanding~\eqref{eq:Zn} gives
\begin{equation}\label{eq:logZn}
\log Z_n = k_g\frac{L}{24 n} + \mathcal{O}(e^{-\frac{L}{2n}})~.
\end{equation}
We then take $L = V_{\mathbb{H}^1} = 2\log(\ell/\epsilon)$ to be infinitely large with $n$ fixed. This suppresses the correction terms in~\eqref{eq:logZn}.
This contributes to the dSRE by
\begin{equation}
\frac{\log Z_n - n\log Z_1}{1 - n} = \frac{k_g}{24}\qty(\frac{1}{n} + 1)V_{\mathbb{H}^1}~,
\end{equation}
which is consistent with gravitational piece of the dSRE in~\eqref{eq:dSRE-6d(1,0)-general}. 
\subsubsection{Holographic dSRE}\label{sec:6d-holo}
We now present a holographic computation of the dSRE for the D4-D2 bound-state defect in 6d $\cN=(1,0)$ SCFTs. As in Section~\ref{sec:4dGW-holography}, the field theory is placed on the replica space $\mathbb{S}^1_n\times\mathbb{H}^5$, with $\beta=2\pi n$, and an R-symmetry background gauge field is turned on to preserve supersymmetry. 
Holographically, this R-symmetry background is realized by the electric charge of a supersymmetric AdS topological black hole.
The corresponding gravitational description is therefore a charged topological black hole in 7d minimal gauged supergravity, uplifted to 10d type \Rm{2}A supergravity via the universal consistent truncation of~\cite{Passias:2015gya}.

The bosonic sector of 7d minimal gauged supergravity consists of the metric, a scalar $X$, an $\SU(2)$ gauge field $A_{(1)}^i$, and a three-form potential.\footnote{In this holographic computation, 7d form fields ($A_{(1)}$, $F_{(2)}$ here and $F_{(4)}$ in Appendix~\ref{appx:D4}) are denoted by italic letters, with the number in parentheses indicating the degree of the differential form.}
For the purely electric Cartan configuration considered here, the three-form can be consistently set to zero, while only the Cartan component $A_{(1)}^3$ is turned on.
The remaining fields are governed by~\cite[Eq.~(2.1)]{Passias:2015gya}\footnote{We have taken $h = g/(2\sqrt{2})$ in~\cite[Eq.~(2.1)]{Passias:2015gya}. }
\begin{equation}
\frac{\mathcal{L}_7}{\sqrt{g}} = R - \frac{1}{2}\qty(\partial\varphi)^2 + g^2\qty(2X^2 + 2X^{-3} - \frac{1}{4}X^{-8}) - \frac{1}{4}X^{-2} F_{(2)\mu\nu} F_{(2)}^{\mu\nu}~,
\end{equation}
where
\begin{equation}
X = e^{\varphi/\sqrt{10}}~,\quad F_{(2)}^i = \dd A_{(1)}^i-\frac{1}{2}g\epsilon^{ijk}A_{(1)}^j\wedge A_{(1)}^k~.
\end{equation}
The solution relevant here can be obtained from the charged AdS$_7$ black hole of~\cite{Cvetic:1999xp} by taking the equal-charge truncation, and is given by\footnote{The relation between the coupling $g$ in~\cite{Passias:2015gya} and the coupling $\mathrm{g}$ in~\cite{Cvetic:1999xp} is $g = \sqrt{2}\mathrm{g}$. }
\begin{equation}\label{eq-7dTBH-main}
\dd s_7^2 = H^{-8/5}f\dd \tau^2 + H^{2/5}(f^{-1}\dd r^2 + r^2 \dd s_{\mathbb{H}^5}^2)~, \quad f = -1 + \frac{1}{8} g^2 r^2 H^2~,\quad H = 1 + \frac{q}{r^4}~.
\end{equation}
The scalar $X$ and electric Cartan gauge field $A_{(1)}$ are
\begin{equation}\label{eq:X-TBH}
    X(r)=H(r)^{-1/5}~, \quad A_{(1)}^3 = A_\tau(r)\dd\tau~.
\end{equation}
The undeformed AdS$_7$ vacuum is recovered when $q=0$, for which $H=X=1$. Using a method similar to that in  Section~\ref{sec:4dGW-holography}, one can obtain the black hole horizon~\cite{Yuan:2023oni, Huang:2025eze}\footnote{We set the AdS$_7$ radius $L_{\rm AdS_7} = 2\sqrt{2}/g = 1$ in the following. }
\begin{equation}\label{eq:7d-horizon}
    r_H = \frac{3n + 1}{4 n}~.
\end{equation}

We now specialize the universal uplift of the solution~\eqref{eq-7dTBH-main} to the massless type \Rm{2}A background with $\mathrm{F}_0=0$, relevant for the class of 6d $\cN=(1,0)$ theories considered here.\footnote{In this holographic calculation, the 10d form fields appearing in the general uplift of~\cite{Passias:2015gya} are denoted by upright letters. } Introducing
\begin{equation}
Z = z(N-z)~,\quad
x = \frac{2z}{N}-1~,\quad
w = X^5(1 - x^2)+x^2~,
\end{equation}
the 10d metric obtained from the universal truncation is
\begin{equation}
ds_{10}^2 = 8\pi X^{-1/2}\sqrt Z \dd s_7^2 + \frac{2\pi X^{5/2}}{\sqrt Z}\left[\dd z^2+ \frac{Z^2}{N^2 w} Ds_{S^2}^2\right]~,
\label{eq:10d-uplift-X}
\end{equation}
where the metric on the covariantized internal two-sphere $S^2$ is given by
\begin{equation}\label{eq:covariant-S2-main}
Ds_{S^2}^2 = \dd\theta^2 + \sin^2\theta \big(\dd\psi - g A^3_{(1)}\big)^2~.
\end{equation}
The dilaton, NSNS potential, and RR fluxes are obtained from the same universal uplift and are collected in Appendix~\ref{appx:DBI}.

The surface defect is described by a probe D4-brane wrapping~\cite{Apruzzi:2024ark}
\begin{equation}\label{eq:D4-embedding-main}
    M_5=\widetilde M_3\times S^2~,
    \quad
    \sigma^a=(\tau,r,\eta,\theta,\psi)~,
    \quad
    z=j~,
\end{equation}
where $\eta$ parametrizes the $\mathbb{H}^1\subset\mathbb{H}^5$ direction along the defect, and the position $z=j$ is inherited from the calibrated D4-D2 embedding in the undeformed background of~\cite{Apruzzi:2024ark}. In Appendix~\ref{appx:D4-EOM}, we verify that this embedding satisfies the equations of motion, while in Appendix~\ref{appx:D4-supersymmetry}, we demonstrate that it preserves supersymmetry.

The expectation value of the surface defect is determined by the Euclidean probe-brane action including the DBI action as well as the WZ term, 
\begin{equation}\label{eq:D4-action-defect} 
-\log\langle D\rangle_n = I_{\mathrm{D4}} = I_{\rm DBI}+I_{\mathrm{WZ}}~. 
\end{equation}
The DBI contribution is
\begin{equation}\label{eq:DBI-main}
    I_{\rm DBI}
    =
    T_4
    \int_{M_5}
    \dd^5\sigma\,
    e^{-\Phi}
    \sqrt{
    \det\qty(g_{\text{D4}}+\mathcal{F})
    }~,
\end{equation}
where $\Phi$ is the 10d dilaton and $\mathcal{F}$ is the gauge-invariant worldvolume flux. 
Evaluating the DBI action~\eqref{eq:DBI-main} on the embedding~\eqref{eq:D4-embedding-main} gives
\begin{equation}\label{eq:D4-action-n-main}
    I_{\text{DBI}}
    =
    -2kj(N-j)n r_H^2 V_{\mathbb{H}^1}~, 
\end{equation}
where $r_H$ is given in~\eqref{eq:7d-horizon}. The details of the DBI computation are given in Appendix~\ref{appx:DBI}.
The classical WZ term is
\begin{equation}\label{eq:classical-D4-WZ}
    I_{\mathrm{WZ}}
    =
    -T_4
    \int_{M_5}
    e^{\mathcal{F}}\wedge P[\mathbf{C}]~,
\end{equation}
which vanishes for the background and embedding considered here. The details of the WZ analysis are presented in Appendix~\ref{appx:WZ}.
The resulting dSRE is
\begin{equation}\label{eq:dSRE-action-relation-main}
    S_n
    = \frac{\log\langle D\rangle_n - n\log\langle D\rangle}{1 - n} = \frac{kj(N-j)}{8}
    \qty(
    \frac{1}{n}+7
    )
    V_{\mathbb{H}^1}~.
\end{equation}
This result precisely reproduces the contribution proportional to $k_I$ in~\eqref{eq-kI-dSRE}. The contribution proportional to the gravitational anomaly $k_g$ is subleading and is not captured by the classical probe-brane couplings considered here. 
\subsection[Wilson surfaces in 6d $\cN=(2,0)$ SCFTs]{\boldmath Wilson surfaces in 6d $\cN=(2,0)$ SCFTs\unboldmath}
\label{sec:6d-BPS-string}
We now test our prescription on half-BPS Wilson surfaces in 6d $\cN=(2,0)$ SCFTs associated with an ADE Lie algebra $\mathfrak g$.
A Wilson surface $\mathcal D_\lambda$ is labelled by the highest weight $\lambda$ of a representation of $\mathfrak g$.
We take its known defect anomaly polynomial~\cite{Wang:2020xkc} as input:
\begin{equation}\label{eq:Wilson-surface-ap}
    \I=
    \frac{1}{2}(\lambda,\lambda)
    \bigl(c_2(L)-c_2(R)\bigr)
    +(\lambda,\rho)
    \bigl(c_2(I)-c_2(F)\bigr)~,
\end{equation}
where $\rho$ is the Weyl vector and the inner product is normalized so that roots have length squared two.
Here $\SU(2)_L\times\SU(2)_R$ describes rotations transverse to the surface, while $\SU(2)_I\times\SU(2)_F$ belongs to the ambient $\SO(5)_R$ symmetry preserved by the defect.
With respect to the $\cN=(1,0)$ subalgebra, $\SU(2)_I$ is the R-symmetry and $\SU(2)_F$ is a flavor symmetry.
In our conventions, the anomaly coefficients are
\begin{equation}
    k_L=-k_R=(\lambda,\lambda)~,\quad
    k_I=-k_F=2(\lambda,\rho)~,\quad
    k_g=0~.
\end{equation}
Denoting the two $\SO(5)_R$ chemical potentials by $\sigma_{1,2}$, the chemical-potential dictionary is
\begin{equation}
\begin{aligned}
    a_I&=-(\sigma_1+\sigma_2)~,&
    a_F&=-(\sigma_1-\sigma_2)~,\\
    a_L&=\omega_2+\omega_3~,&
    a_R&=\omega_2-\omega_3~.
\end{aligned}
\end{equation}
Using the equivariant parameters $c_2(A) \to -a_A^2/4$ for the R-symmetry background gauge fields, with $A=L,R,I,F$, and choosing $\omega_\ell=\omega_1$, the prescription~\eqref{eq:dSCE-6d(1,0)-general} gives
\begin{equation}\label{eq:dSCE-Wilson-surface}
    E_C =\int_{\mathbb R^2_{\omega_1}}=-\frac{1}{2\omega_1}
    \bigl[
        (\lambda,\lambda)\omega_2\omega_3
        +2(\lambda,\rho)\sigma_1\sigma_2
    \bigr]~,
\end{equation}
which reproduces the Wilson-surface dSCE obtained
in~\cite{Huang:2025eze}.

For the supersymmetric R\'enyi background, the $\cN=(1,0)$ specialization considered above turns on an $n$-dependent background field only for the $\SU(2)_I$ Cartan. 
In the notation of~\cite{Huang:2025eze}, this corresponds to $r_1=r_2=\tfrac12$, so that $a_F$ has no $n$ dependence.
Using $k_I=2(\lambda,\rho)$ and $k_g=0$ in~\eqref{eq:dSRE-6d(1,0)-general}, we obtain the universal contribution
\begin{equation}\label{eq:dSRE-Wilson-surface}
    S_n=
    \frac{(\lambda,\rho)}{4}
    \left(\frac{1}{n}+7\right)V_{\mathbb H^1}~.
\end{equation}
This agrees with the same specialization of the $\cN=(2,0)$ result in~\cite{Huang:2025eze} and reproduces the $\SU(2)_I$ universal piece~\eqref{eq-kI-dSRE}.

\section{Discussion}
\label{sec:discussion}

We have extended the anomaly-polynomial prescriptions for dSCE and dSRE to Gukov--Witten defects in four-dimensional $\mathcal{N} = 4$ SYM and half-BPS surface defects in six-dimensional $\mathcal{N} = (1,0)$ SCFTs. 
On the supersymmetric backgrounds considered here, the results are governed by defect 't~Hooft anomalies, with an explicit gravitational-anomaly contribution for chiral defects. 
Complementary field-theory and holographic calculations test different terms and limits of these expressions. 
These results support anomaly control of protected observables even when the defect degrees of freedom remain coupled to the ambient theory.

A concrete next step is to reproduce holographically the contributions
\begin{equation}
  E_C\supset-\frac{k_g\omega_1}{24}~,
  \quad
  S_n\supset
  \frac{k_g}{24}\left(1+\frac1n\right)V_{\mathbb H^1}~.
\end{equation}
The curvature coupling in the D4-brane Wess--Zumino action induces a three-dimensional gravitational Chern--Simons term whose coefficient reproduces $k_g$~\cite{Apruzzi:2024ark}.
Matching this anomaly coefficient does not by itself establish the corresponding supersymmetric energy and entropy. 
It remains to determine whether the relevant supersymmetric completion and boundary terms, or additional higher-derivative corrections, reproduce the expressions above. 
This would provide a test beyond the leading probe calculation and clarify the mechanism underlying the gravitational-anomaly contributions.

A remaining question is how localization captures the gravitational-anomaly contribution $E_C^{(g)}=-k_g\omega_1/24$.
Our large-$N$ Wilson-loop calculation reproduces the leading $k_I$ contribution, but does not resolve this subleading term.
Its extraction requires a more precise identification of the five-dimensional operator representing the surface defect, including any localized degrees of freedom. It remains to determine whether the $k_g$ term arises from corrections to the Wilson-loop expectation value or requires additional defect contributions.

A reliable extraction of $k_g$ requires a more microscopic characterization of the defect operator, for example from an explicit brane construction or an equivalent derivation of the coupled lower-dimensional defect theory. We leave this question for future investigation.

Another question concerns Gukov--Witten defects with nonzero $\beta_i, \gamma_i$. 
These moduli break the independent transverse $\SO(2)_\perp$ rotation while preserving a diagonal combination with R-symmetry. 
The defect Weyl anomaly $d_2$ then acquires continuous moduli dependence. 
Extending our analysis therefore requires identifying the compatible supersymmetric backgrounds and their chemical-potential constraints. 
One can then ask whether the defect anomaly polynomial still determines the observables completely, or whether additional moduli-dependent data enter. 

A field-theoretic derivation of the extended equivariant prescription would also be valuable. 
One possible route is to study anomalous modular transformations of suitable protected partition functions and their relation to the low-temperature limits defining the dSCE. 
Such an approach must account for coupling to the ambient theory rather than assume an autonomous two-dimensional CFT. 

Further tests include surface defects with different types of defects~\cite{Faedo:2020nol, Capuozzo:2024onf}. 
A distinct extension is to codimension-two defects in six-dimensional SCFTs. 
These four-dimensional defects are characterized by a six-form anomaly polynomial $I^{(6)}$, providing an opportunity to test the prescription beyond two-dimensional defects.

\acknowledgments
We are grateful for useful discussions with our group members at Fudan University. This work is supported by NSFC Grant No. 12375063. This work is also sponsored by Shanghai Oriental Talents Program. YZ is also supported by NSFC Grant No. 12247103 through Peng Huanwu Center for Fundamental Theory.
\appendix
\section{Conventions and equivariant integral}
In this appendix, 
we summarize our conventions for characteristic classes and the equivariant integral used in the main text. 

For an $\SU(2)$ bundle with the Cartan generator~$J$ and background gauge field strength~$F$,
we normalize the $J$ in the fundamental representation as
\begin{equation}
    J=\frac12\operatorname{diag}(1,-1)~,
\qquad
\text{tr}_{\mathrm{fund}}J^2=\frac12~.
\end{equation}
We also absorb the factor \(1/(2\pi)\) into the definition of the field strength so that the total Chern class is $c(E)=\det(1+iF)$. Thus the second Chern class is defined as
\label{appx:conv}
\begin{equation}
c_2(F)=\frac{1}{2}\text{tr}_{\text{fund}} F^2\equiv \frac{1}{2}\tr F^2~.
\end{equation}
For $\SO(4)\simeq \SU(2)_{\ttL}\times\SU(2)_{\ttR}$, the corresponding Euler class is
\begin{equation}
\chi(\SO(4))=\frac{1}{2}\qty(\text{tr}_{\text{fund}} F_\ttL^2-\text{tr}_{\text{fund}} F_\ttR^2)=c_2(\ttL)-c_2(\ttR)~.
\end{equation}
We use the Duistermaat-Heckman (DH) formula to evaluate the equivariant integral of the defect anomaly polynomial. The DH formula relates the equivariant integral of a differential form $I$ 
on $\mathbb{R}^{d}$ to the value of the $0$-form component of $I$ at the origin, 
\begin{equation}
    \int_{\mathbb{R}^{d}} I =  \frac{I |_0}{e(TM)|_0}~,
\end{equation}
where $I |_0$ and $e(TM)|_0$ are the 0-form component of $I $ and the equivariant Euler class at the origin, respectively. 
We now illustrate how to use the DH formula to evaluate a generic anomaly $4$-form on $\mathbb{R}^2_{\omega}$,
\begin{equation}
    \I=\frac{1}{2}k\,c_2(F)-\frac{1}{24}k_g\, p_1(TM)~.
\end{equation}
At the origin, the $0$-form component of the equivariant characteristic classes $c_2(F)$, $p_1(TM)$ and $e(TM)$ are given by the equivariant parameters,
\begin{equation}
    c_2(F)|_0 =-\frac{a^2}{4}~,\quad p_1(TM)\to \omega^2~, \quad e(TM)|_0 = \omega~.
\end{equation}
The DH formula then gives
\begin{equation}
    \int_{\mathbb{R}^{2}_\omega} \I =-\frac{1}{8 \omega}k\,a^2 -\frac{1}{24} k_g \, \omega~.
\end{equation}
\section{Chemical potentials in the Schur limit}
\label{app:Schur-chemical}
First we need the relation between the 4d $\cN=2$ chemical potentials and the $\cN=4$ ones. This can be done by comparing the expressions of the corresponding superconformal indices. The 4d $\cN=2$ superconformal index that preserves the supercharge $Q^{R,r}_{J_1,J_2}=Q^{+-}_{--}$ with commutator $\{Q,Q^\dagger\}=E -J_1-J_2-2R+r$ is typically written as
\begin{equation}
    \cI_{\cN =2} = \tr (-1)^F p^{J_1-r} q^{J_2-r} t^{r+R} a^f~,
\end{equation}
with the R-symmetry $\SU(2)_R \times\U(1)_r$, flavor symmetry $G_f$ and the fugacities
\begin{equation}
    p=e^{-\beta \omega_1}~, \quad q=e^{-\beta \omega_2}~,\quad t=e^{-\beta \mu}~,\quad a=e^{-\beta m}\ .
\end{equation} 
Meanwhile, the 4d $\cN=4$ superconformal index
\begin{equation}
    \cI_{\cN=4} = \tr\qty[ (-1)^F e^{-\beta\left(\sum_{i=1}^2\omega_iJ_i+\sum_{I=1}^3 \Delta_I R_I \right)}]
\end{equation}
with the BPS condition 
\begin{equation} 
\label{eq:4d-BPS-cdt}\sum_{i=1}^2\omega_i+\sum_{I=1}^3 \Delta_I=0
\end{equation}
preserves the supercharge $Q$ with the eigenvalues $(E, J_1, J_2, R_1, R_2, R_3) = (\frac{1}{2}, -\frac{1}{2}, -\frac{1}{2}, -\frac{1}{2}, -\frac{1}{2}, -\frac{1}{2})$ and the commutator $\{Q,Q^\dagger\} = E -J_1 - J_2 + R_1 + R_2 + R_3$.
The Cartans of 4d $\cN=2$ subalgebra are embedded in the $\cN=4$ ones as
\begin{equation}
    R=-\frac{1}{2}(R_1+R_2)~, \quad r=R_3~, \quad f=\frac{1}{2}(R_1-R_2)~,
\end{equation}
with the chemical potentials
\begin{equation}
    \mu=-\Delta_1-\Delta_2~, \quad m= \Delta_1-\Delta_2\ .
\end{equation}
The Schur limit $q=t$ gives $\omega_2=\mu$, which combined with the BPS condition~\eqref{eq:4d-BPS-cdt} leads to $\Delta_3=-\omega_1$. The flavor fugacity is turned off, $m=0$, as the setting in~\cite{Alday:2013kda}. 
Combining all these conditions gives the Schur limit chemical potential setting,
\begin{equation}
\label{eq:chem-Schur}
    \Delta_1=\Delta_2=-\frac{1}{2}\omega_2~,\quad \Delta_3=-\omega_1~.
\end{equation}
\section{\boldmath Gukov--Witten defect anomaly from 2d GLSM \unboldmath}
\label{appx:GW-anomaly-polynomial}
In this section we show the derivation of defect anomaly polynomial from the field content of the 2d GLSM description of the Gukov--Witten defect.
We use the notation that a 2d Weyl fermion transforming in a representation $\mathcal{R}$ of a symmetry group $G$ (including both the gauge and global symmetry), with the chirality labeled by $\pm$, has the anomaly four-form
\begin{equation}
   \pm\qty[\hat{a}(T\mathcal{M})\cdot \str_\mathcal{R}(e^{F_G}) ]_4~,
\end{equation}
where $\hat{a}(T\mathcal{M})$ is the A-roof genus corresponding to the 2d spacetime manifold $\mathcal{M}$, $F_G$ is the background gauge field strength for the group $G$.
The 2d GLSM has global symmetry $\SU (2)_\ttL\times\SU(2)_\ttR\times\SU(2)_\ttF$~\cite{Wang:2020xkc}. The Cartan generators of $\mathrm{SU}(2)_{\ttL,\ttR}$ are identified with the bulk generators as in~\eqref{eq-cartan-id1}, while the Cartan generator of $\mathrm{SU}(2)_\ttF$ is identified as $J_\ttF=R_1$~\cite[Appendix C]{Chen:2023lzq}.
\subsection{Simple defect}
In this case, the 2d GLSM is simply a $(4,4)$ SQED with only one single $\U(1)$ gauge node and a $\SU(N)$ flavor node. The field content consists of a $(4,4)$ vector multiplet as well as $N$ hypermultiplets. 
The gauginos $\lambda_\pm$ and the hypermultiplet fermions $\psi_\pm$ transform in the $\SU(2)_\ttL\times\SU(2)_\ttR$ R-symmetry and $\SU(2)_\ttF$ flavor symmetry as shown in~\eqref{eq:GLSM-charge}. 
The Cartan charges are shown in Table~\ref{tab:GLSM-fermion-charge}. We normalize the charges to be half-integers, $\pm\frac{1}{2}$. Notice that the hypermultiplet fermions also transform in the fundamental representation of the gauge $\U(1)$ symmetry.
\begin{table}[!htbp]
\centering
\begin{tabular}{|l|lll|}
\hline
 & $J_\ttL$ & $J_\ttR$ & $J_\ttF$ \\ \hline
$\lambda_+$ & $-\frac{1}{2}$ & $0$ & $-\frac{1}{2}$ \\
$\lambda_-$ & $0$ & $-\frac{1}{2}$ & $-\frac{1}{2}$ \\
$\psi_+$ & $0$ & $\frac{1}{2}$ & $0$ \\
$\psi_-$ & $\frac{1}{2}$ & $0$ & $0$ \\ \hline
\end{tabular}
\caption{The Cartan charges of gauginos and hypermultiplet fermions in the global symmetry $\SU(2)_\ttL \times\SU(2)_\ttR \times\SU(2)_\ttF$. We omit the conjugate fermions $\bar{\lambda}, \bar{\psi}$; their charges are opposite to those of $\lambda, \psi$, respectively. We follow the convention of Appendix C in~\cite{Chen:2023lzq}.}
\label{tab:GLSM-fermion-charge}
\end{table}
\FloatBarrier
The fermions contribute to the anomaly polynomial as
\begin{equation}
    \begin{aligned}
        I_{(\lambda_+)}
        &=-\frac{1}{12}p_1(T)+\frac{1}{2}\str_{\text{fund}}F_\ttL^2+\frac{1}{2}\str_{\text{fund}}F_f^2+2\big[\str_{\text{adj}}e^{F_{\U(1)}}\big]_4~,~\\
        I_{(\lambda_-)}
        &=-\frac{1}{12}p_1(T)+\frac{1}{2}\str_{\text{fund}}F_\ttR^2+\frac{1}{2}\str_{\text{fund}}F_f^2+2\big[\str_{\text{adj}}e^{F_{\U(1)}}\big]_4~,\\
        I_{(\psi_+)}
        &=-\frac{1}{12}p_1(T)+\frac{1}{2}\str_{\text{fund}}F_\ttR^2+2\big[\str_{\text{fund}}e^{F_{\U(1)}}\big]_4~,\\
        I_{(\psi_-)}
        &=-\frac{1}{12}p_1(T)+\frac{1}{2}\str_{\text{fund}}F_\ttL^2+2\big[\str_{\text{fund}}e^{F_{\U(1)}}\big]_4~,\\
    \end{aligned}    
\end{equation}
where $F_{\ttL,\ttR,f}$ are the background field strength associated with $\SU(2)_{\ttL,\ttR,\ttF}$, and $F_{\U(1)}$ is that of the gauge $\U(1)$ group. 
The total defect anomaly is given by summing over all the fermion contribution,
\begin{equation}
 \I = I_{(\lambda_{+})} - I_{(\lambda_{-})} + N\qty(I_{(\psi_+)} - I_{(\psi_-)}) = -\frac{1}{2}(N-1) \qty(\str_{\text{fund}} F_\ttL^2-\str_{\text{fund}} F_\ttR^2)~.
\end{equation}
\subsection{General defect}\label{apdx:GW-field}
Now we turn to Gukov--Witten defect labeled by a general Levi subgroup $\bL=\S[\prod_{i=1}^{\nu}\U(N_i)]$.
The 2d $(4,4)$ GLSM is a linear quiver theory
\[
[\SU(p_\nu)] \times \U(p_{\nu - 1}) \times \cdots\times\U(p_{1})
\]
with the field content
\begin{itemize}
\item One $(4,4)$ vector multiplet for each gauge node $\U(p_i)$, composed of one $(2,2)$ vector multiplet $V$ and one $(2,2)$ chiral multiplet $X$, both in the adjoint representation of $\U(p_i)$.
\item One $(4,4)$ hyper multiplet for each $\U(p_i)\times\U(p_{i+1})$, composed of one $(2,2)$ bifundamental chiral multiplet $\Phi_{i(i+1)}$ transformed in $(\bold{p_i}, \bold{\bar{{p}}_{i+1}})$ and one $(2,2)$ anti-bifundamental chiral multiplet $\Phi_{(i+1)i}$ transformed in $(\bold{\bar{{p}}_i},\bold{p_{i+1}})$. At the end of the gauge node, $\Phi_{(\nu-1)\nu}$ and $\Phi_{\nu(\nu-1)}$ are transformed in $\bold{p_{\nu-1}}$ and $\bold{\bar{{p}}_{\nu-1}}$, respectively, but both transformed in the fundamental representation of $\SU(N)$.\footnote{In this sense it is also convenient to state that there are $N$ fundamental hypers for $\U(p_{\nu-1})$}
\end{itemize}
For the $i$-th gauge node with gauge symmetry $\U(p_i)$, the fermions contribute to the anomaly polynomial as
\begin{equation}
\begin{aligned}
I_{\lambda^{(i)}_+}
&= p_i^2\qty(-\frac{1}{12}p_1(T) + \frac{1}{2}\str_{\text{fund}}F_\ttL^2 + \frac{1}{2} \str_{\text{fund}}F_f^2) + 2\big[\str_{\text{adj}} e^{F_{\U(p_i)}}\big]_4~,\\
I_{\lambda^{(i)}_-}
&= p_i^2\qty(-\frac{1}{12}p_1(T)+\frac{1}{2} \str_{\text{fund}}F_\ttR^2 + \frac{1}{2} \str_{\text{fund}} F_f^2) + 2\big[\str_{\text{adj}} e^{F_{\U(p_i)}}\big]_4~,\\    
I_{\psi^{ (i)}_+}
&=p_i p_{i+1}\qty(-\frac{1}{12}p_1(T) + \frac{1}{2} \str_{\text{fund}} F_\ttR^2) + 2\big[\str_{\text{fund}} e^{F_{\U(p_i)}}\cdot \str_{\text{fund}}e^{F_{\U(p_{i+1})}}\big]_4~, \\
I_{\psi^{(i)}_-} 
&= p_i p_{i+1}\qty(-\frac{1}{12}p_1(T) + \frac{1}{2} \str_{\text{fund}} F_\ttL^2) + 2\big[\str_{\text{fund}} e^{F_{\U(p_i)}}\cdot \str_{\text{fund}} e^{F_{\U(p_{i+1})}}\big]_4~,
\end{aligned}
\end{equation}
with $p_i^2$ in $I_{\lambda_{\pm}^{(i)}}$ coming from the trace over adjoint representation of gauge $\U(p_i)$, $\str_{\text{adj}} \mathds{1}_{\U(p_i)}$, $p_i p_{i+1}$ in $I_{\psi_{\pm}^{(i)}}$ coming from the trace over bi-fundamental representation of $\U(p_i)\times\U(p_{i+1})$.
For the last gauge node, the hypermultiplet fermions are in the bi-fundamental representation of $\U(p_{\nu-1})\times\SU(N)$,
\begin{equation}
    \begin{aligned}
        I_{\psi^{(\nu-1)}_+} 
        &=p_{\nu-1} p_{\nu}\qty(
    -\frac{1}{12}p_1(T)+\frac{1}{2}\str_{\text{fund}}F_\ttR^2)+2\big[\str_{\text{fund}}e^{F_{\U(p_{\nu-1})}}\cdot \str_{\text{fund}}e^{F_{\SU(p_{\nu})}}\big]_4~, \\
    I_{\psi^{(\nu-1)}_-}
    &=p_{\nu-1} p_{\nu}\qty(
    -\frac{1}{12}p_1(T)+\frac{1}{2}\str_{\text{fund}}F_\ttL^2)+2\big[\str_{\text{fund}}e^{F_{\U(p_{\nu-1})}}\cdot \str_{\text{fund}}e^{F_{\SU(p_{\nu})}}\big]_4~,
    \end{aligned}
\end{equation}
where we have defined $p_\nu=N$. Summing over all the fermions leads to the defect anomaly polynomial~\eqref{eq:GW-ap-general},
\begin{equation}
\begin{split}
\I
&=\sum_{i=1}^{\nu-1}I_{\lambda_{+}^{(i)}}-I_{\lambda_{-}^{(i)}}+I_{\psi_{+}^{(i)}}-I_{\psi_{-}^{(i)}}\\
&= -\frac{1}{2}\sum_{i=1}^{\nu-1}\qty(p_ip_{i+1}-p_i^2)\qty(\str_{\text{fund}}F_\ttL^2-\str_{\text{fund}}F_\ttR^2)~.
\end{split}
\end{equation}
\section{Probe D4-brane action}\label{appx:D4}
In this appendix, we provide the details of the probe D4-brane calculation in Section~\ref{sec:6d-holo}. 
We first establish the dictionary between the conventions of~\cite{Apruzzi:2024ark} and~\cite{Passias:2015gya} and collect the 10d supergravity fields required to evaluate the DBI and WZ actions.

The undeformed $F_0=0$ type \Rm{2}A $\mathrm{AdS}_7\times M_3$ solution in~\cite{Apruzzi:2024ark} is
\begin{align}
\frac{1}{\pi \sqrt{2}}\dd s^2 &= 8\sqrt{-\frac{\alpha}{\alpha''}}\dd s_{\mathrm{AdS}_7}^2 + \sqrt{-\frac{\alpha''}{\alpha}}\qty(\dd z^2 + \frac{\alpha^2}{\alpha'^2 - 2\alpha \alpha''}\dd s_{\mathbb{S}^2}^2)~,\\
e^\phi &= 3^4 (\sqrt{2}\pi)^{5/2}\frac{(-\alpha/\alpha'')^{3/4}}{\sqrt{\alpha'^2 - 2\alpha\alpha''}}~,\label{eq:undeform-phi}\\
B &= \pi\qty(-z + \frac{\alpha\alpha'}{\alpha'^2 - 2\alpha\alpha''})\mathrm{vol}_{\mathbb{S}^2}~,\label{eq:undeform-B}\\
F_2 &= \qty(\frac{\alpha''}{162\pi^2} + \frac{\pi F_0 \alpha \alpha'}{\alpha'^2 - 2\alpha\alpha''})\mathrm{vol}_{\mathbb{S}^2}~,
\end{align}
where $\dd s_{\mathrm{AdS}_7}^2$ and $\dd s_{\mathbb{S}^2}^2$ denote the line elements on AdS$_7$ and the unit two-sphere $\mathbb{S}^2$, respectively, and a prime denotes differentiation with respect to $z$. For the $F_0=0$ solution considered here,
\begin{equation}
    \alpha(z)=\frac{81}{2}\pi^2 k z(N-z)~.
\end{equation}

The general 10d uplift ansatz with nontrivial $X$ is~\cite[Section~4.1]{Passias:2015gya}\footnote{We have fixed a typo in Eq.~(4.7) of~\cite{Passias:2015gya}: in~\eqref{eq:general-B}, $A_{(1)}^i$ is replaced by $gA_{(1)}^i$. }
\begin{align}
\ell^{-1}\dd s_{10}^2 =&\, \frac{1}{8}g^2 X^{-1/2}e^{2A}\dd s_7^2 + X^{5/2}\dd s_{M_3}^2~,\quad \dd s_{M_3}^2 = \dd \rho^2 + \frac{1 - x^2}{16 w}e^{2A} D s_{S^2}^2~,\label{eq:uplift-metric}\\
e^{2\Phi} =&\, \ell \frac{X^{5/2}}{w}e^{2\tilde{\phi}}~,\label{eq:general-phi}\\
\ell^{-1} \mathrm{B} =&\, \frac{1}{16}e^{2A} \frac{x\sqrt{1 - x^2}}{w}\mathrm{vol}_{S^2} - \frac{1}{2}e^A \dd\rho\wedge\qty(a - \frac{1}{2}y^i g A_{(1)}^i)~,\label{eq:general-B}\\
\mathrm{F}_2 =&\, -\mathfrak{q}\qty(\mathrm{vol}_{S^2} + y^i g F_{(2)}^i) + \frac{1}{16}w^{-1}\ell\, \mathrm{F}_0 e^{2A} x\sqrt{1 - x^2}\mathrm{vol}_{S^2}~,\\
\ell^{-1}\mathrm{F}_4 =&\, -\frac{\mathfrak{q}}{16}e^{2A}\frac{x\sqrt{1-x^2}}{w}y^i g F_{(2)}^i\wedge\mathrm{vol}_{S^2} - \frac{\mathfrak{q}}{4} e^A \dd\rho \wedge \epsilon^{ijk}g F_{(2)}^i \wedge y^j D y^k\\
&\,-\frac{\mathfrak{q}}{2}e^A\dd\rho\wedge X^4 g^2 *_7 F_{(4)} - \ell^{-1}\frac{1}{2}e^{3A - \tilde{\phi}}x F_{(4)}~.\label{eq:general-F4-appx}
\end{align}
where $\ell = 8\sqrt{2}/g^3$, $\tilde{\phi}$ is the dilaton for the undeformed $\mathrm{AdS}_7\times M_3$ background, $a$ is defined via $\dd a = -\frac{1}{2}\mathrm{vol}_{\mathbb{S}^2} = -\frac{1}{4}\epsilon^{ijk}y^i \dd y^{j}\wedge \dd y^{k}$, $\mathfrak{q} = \frac{1}{4}e^{A - \tilde{\phi}}\sqrt{1 - x^2}$, $F_{(2)}^i = \dd A_{(1)}^i - \frac{1}{2}g\epsilon^{ijk}A_{(1)}^j\wedge A_{(1)}^k$, $F_{(4)}$ is the field strength of the 7d three-form and vanishes in our setup here, and $w = X^5(1 - x^2) + x^2$. The volume of the covariantized $S^2$ is $\mathrm{vol}_{S^2} = \frac{1}{2}\epsilon^{ijk}y^i D y^{j}\wedge D y^{k}$ and the covariantized metric $Ds_{S^2}^2$ on the two-sphere is
\begin{equation}
\quad Ds_{S^2}^2 = Dy^i Dy^i~,\quad Dy^i\equiv \dd y^i + \epsilon^{ijk}y^j g A_{(1)}^k~,\quad y^i = (\sin\theta\cos\psi, \sin\theta\sin\psi, \cos\theta)~. 
\end{equation}
For the Cartan background considered here, only
$A_{(1)}^3=A_\tau(r)\dd\tau$ is turned on, and the covariantized metric reduces to~\eqref{eq:covariant-S2-main}.

We now specialize the universal uplift ansatz to the $\mathrm{F}_0=0$ background introduced above. 
Introducing the shorthand $Z=z(N-z)$, the dictionary between the parametrization of~\cite{Apruzzi:2024ark} and the variables used in the universal uplift is
\begin{equation}\label{eq:match0}
x = \frac{2z}{N} - 1~,\quad 1-x^2=\frac{4Z}{N^2}~.
\end{equation}
Matching the uplift ansatz to the undeformed solution fixes
\begin{equation}\label{eq:match}
g=2\sqrt2~,\quad \ell=\frac{1}{2}~,\quad e^{2A}=16\pi\sqrt{Z}~,\quad \dd\rho = -2\sqrt{\pi} Z^{-1/4}\dd z~,\quad e^{2\tilde{\phi}} = \ell^{-1} e^{2\phi}~,
\end{equation}
and the Neveu-Schwarz potential $\mathrm{B}$ in~\eqref{eq:general-B} differs from that in~\eqref{eq:undeform-B} by a minus sign and a gauge transformation when $A_{(1)}^i = 0$,
\begin{equation}
-\mathrm{B} = B - \dd (2\pi z a)~.
\end{equation}

Substituting~\eqref{eq:match0} and \eqref{eq:match} into~\eqref{eq:uplift-metric} gives~\eqref{eq:10d-uplift-X}:
\begin{equation}
ds_{10}^2 = 8\pi X^{-1/2}\sqrt Z \dd s_7^2 + \frac{2\pi X^{5/2}}{\sqrt Z}\left[\dd z^2+ \frac{Z^2}{N^2 w} Ds_{S^2}^2\right]~,
\label{eq:10d-uplift-X-appx}
\end{equation}
with the 7d metric
\begin{equation}\label{eq-7dTBH}
\dd s_7^2 = H^{-8/5}f\dd \tau^2 + H^{2/5}(f^{-1}\dd r^2 + r^2 \dd s_{\mathbb{H}^5}^2)~, \quad f = -1 + r^2 H^2~,\quad H = 1 + \frac{q}{r^4}~,
\end{equation}
and the scalar $X(r) = H(r)^{-1/5}$. In~\eqref{eq-7dTBH} we have already set $g = 2\sqrt{2}$.
\subsection{Dirac-Born-Infeld action}\label{appx:DBI}
We now evaluate the DBI action~\eqref{eq:DBI-main} in the background~\eqref{eq:10d-uplift-X-appx} with the D4-brane embedding~\eqref{eq:D4-embedding-main}. For the embedding at $z=j$, it is convenient to define
\begin{equation}
    Z_j=j(N-j)~,\quad
    Y_j(r) = N^2 w|_{z = j} = (N-2j)^2 + 4X^5Z_j~.
\end{equation}
On the D4-brane, the warp factors multiplying the external and internal parts of the induced metric are, respectively,
\begin{equation}\label{eq:wrap}
\mathcal{W}_7(r) = 8\pi X(r)^{-1/2}\sqrt{Z_j}~,\quad \mathcal{W}_3(r) = \frac{2\pi X(r)^{5/2}Z_j^{3/2}}{Y_j(r)}~.
\end{equation}
The induced metric on the D4-brane therefore takes the form
\begin{equation}
g_{\text{D4}} = \mathcal{W}_7\, \dd s_{\widetilde M_3}^2 + \mathcal{W}_3 \left[\dd\theta^2 + \sin^2\theta(\dd\psi - gA_\tau \dd\tau)^2\right]~.
\end{equation}

Using the undeformed dilaton~\eqref{eq:undeform-phi}, the uplift formula~\eqref{eq:general-phi}, and the dictionary~\eqref{eq:match}, we obtain
\begin{equation}
e^\Phi = \frac{2\sqrt{2\pi}}{k} \frac{X^{5/4}Z^{3/4}}{\sqrt{(N-2z)^2+4X^5Z}}~.
\end{equation}
Restricting to $z=j$, this becomes
\begin{equation}\label{eq:phi}
e^{-\Phi_j} = \frac{k}{2\sqrt{2\pi}} X^{-5/4}Z_j^{-3/4}\sqrt{Y_j}~.
\end{equation}

We next determine the gauge-invariant worldvolume two-form $\mathcal{F}$. 
In the conventions of~\cite{Apruzzi:2024ark}, $\mathcal{F} = 2\pi\hat{f} + P[B]$, whereas in the uplift conventions adopted here we use $\mathcal{F} = 2\pi\hat{\mathrm{f}}-P[\mathrm B]$. 
Since $B$ and $(-\mathrm{B})$ differ by a gauge transformation, the worldvolume gauge field must be transformed accordingly.
When the $\SU(2)$ gauge field is turned on, the worldvolume flux of~\cite{Apruzzi:2024ark} admits the gauge-covariant extension
\begin{equation}\label{eq:flux}
\hat{f} = \frac{j}{2}\mathrm{vol}_{\mathbb{S}^2} \to \hat{f} = \frac{j}{2}\qty(\mathrm{vol}_{S^2} + y^i g F_{(2)}^i)~.
\end{equation}
Indeed,
\begin{equation}
\dd\qty(
a-\frac{1}{2}y^igA_{(1)}^i
)
=
-\frac{1}{2}
\qty(
\mathrm{vol}_{S^2}+y^igF_{(2)}^i
)~,
\end{equation}
which immediately implies $\dd\hat f=0$. Under the corresponding gauge transformation,
\begin{equation}\label{eq:gauge-transf}
2\pi\hat{\mathrm{f}} = 2\pi\hat{f} + \dd \Lambda~,\quad \Lambda = 2\pi z\qty(a - \frac{1}{2}y^i g A_{(1)}^i)~.
\end{equation}
Upon pullback to the D4-brane at $z=j$, one finds
\begin{equation}\label{eq:f-0}
2\pi\hat{\mathrm{f}} = \pi j\qty(\mathrm{vol}_{S^2} + y^i g F_{(2)}^i) + \dd \Lambda|_{z = j} = 0~.
\end{equation}
Therefore, in this gauge, the gauge-invariant worldvolume flux is entirely determined by the pullback of the NSNS potential. Substituting~\eqref{eq:match} to~\eqref{eq:general-B} gives
\begin{equation}\label{eq:flux-new}
\mathcal{F} = 2\pi\hat{\mathrm{f}} - P[\mathrm{B}] = \pi\frac{Z_j(N-2j)}{Y_j(r)}\mathrm{vol}_{S^2} \equiv \varphi(r)\,\mathrm{vol}_{S^2}~,\quad \mathrm{vol}_{S^2} = \sin\theta\,\dd\theta\wedge(\dd\psi - gA_\tau \dd\tau)~. 
\end{equation}
As a consistency check, in the undeformed limit $X=1$ this reproduces the flux of~\cite{Apruzzi:2024ark},
\begin{equation}
\varphi\big|_{X=1} = \pi\frac{Z_j(N - 2j)}{N^2} = \pi\frac{\alpha\alpha'}{\alpha'^2 - 2\alpha\alpha''}\Big|_{z = j}~.
\end{equation}

Now we compute the DBI determinant. Let
\begin{equation}
    \mathrm{s} \equiv \sin\theta~,\quad \mathrm{a}(r)\equiv gA_\tau(r)~.
\end{equation}
For compactness, we introduce the following components of the induced metric in the coordinate basis $(\tau,r,\eta,\theta,\psi)$,
\begin{equation}
    G_\tau = \mathcal{W}_7 H^{-8/5}f~,\quad
    G_r = \mathcal{W}_7 H^{2/5}\frac{1}{f}~,\quad
    G_\eta = \mathcal{W}_7 H^{2/5}r^2~,\quad
    G_\theta  = \mathcal{W}_3~.
\end{equation}
Then the matrix entering the DBI action is 
\begin{equation}
[g_{\text{D4}}]_{ab}+\mathcal{F}_{ab}
=
\begin{pmatrix}
G_\tau+G_\theta\,\mathrm{s}^2\,\mathrm{a}^2 & 0 & 0 & +\varphi\,\mathrm{s}\,\mathrm{a} & -G_\theta\,\mathrm{s}^2\mathrm{a}\\
0 & G_r & 0 & 0 & 0\\
0 & 0 & G_\eta & 0 & 0\\
-\varphi\,\mathrm{s}\,\mathrm{a} & 0 & 0 & G_\theta  & +\varphi\,\mathrm{s}\\
-G_\theta\,\mathrm{s}^2\mathrm{a} & 0 & 0 & -\varphi\,\mathrm{s} & G_\theta\,\mathrm{s}^2
\end{pmatrix}~.
\label{eq:D4-matrix}
\end{equation}
Its determinant simplifies to
\begin{equation}
\det\qty(g_{\mathrm{D4}}+\mathcal{F})
=
G_\tau G_r G_\eta \sin^2\theta \qty(G_\theta^2 + \varphi^2)~,
\end{equation}
so that the explicit dependence on $A_\tau$ cancels.
The DBI action is then given by 
\begin{equation}\label{eq:DBI}
\begin{split}
I_{\rm DBI} 
&= T_4\int_{M_5} \dd^5\sigma\, e^{-\Phi_j}\sqrt{\det \qty(g_{\text{D4}} + \mathcal{F})}\\
&= \frac{1}{16\pi^4}\int_{\mathbb{S}_n^1} \!\dd\tau \int_{\mathbb{H}^1} \!\dd\eta \int_{r_H}^\Lambda \!\dd r \int_0^\pi\! \dd\theta \int_0^{2\pi} \!\!\dd\psi~e^{-\Phi_j} \mathcal{W}_7^{3/2}H^{-2/5} \sqrt{\mathcal{W}_3^2+\varphi^2}\, r\sin\theta~,
\end{split}
\end{equation}
where we have used $T_4 = 1/(16\pi^4 l_s^5)$ with $l_s = 1$. Substituting~\eqref{eq:wrap}, \eqref{eq:phi}, \eqref{eq:flux-new}, and $X=H^{-1/5}$ into~\eqref{eq:DBI}, we obtain
\begin{equation}
\begin{split}
I_{\text{DBI}} &= \frac{1}{16\pi^4}\int_{\mathbb{S}_n^1} \!\dd\tau \int_{\mathbb{H}^1} \!\dd\eta \int_{r_H}^\Lambda \!\dd r \int_0^\pi\! \dd\theta \int_0^{2\pi} \!\!\dd\psi~8\pi^2 k Z_j r \sin\theta\\
&= 2 k j(N - j)n\qty(\Lambda^2 - r_H^2)V_{\mathbb{H}^1}~.
\end{split}
\end{equation}
After subtracting the power-law divergence at the asymptotic AdS boundary, we obtain the renormalized DBI action~\eqref{eq:D4-action-n-main}. 
\subsection{Wess--Zumino term}\label{appx:WZ}
It is convenient to evaluate the WZ term in the alternative basis of RR potentials introduced in~\cite{Bergshoeff:2001pv},\footnote{Boldface symbols denote formal sums. In~\eqref{eq:A-basis-definition}, 
\[
\mathbf{C} = \mathrm{C}_1 + \mathrm{C}_3 + \mathrm{C}_5 + \mathrm{C}_7 + \mathrm{C}_9~, \quad \mathbf{A} = \mathrm{A}_1 + \mathrm{A}_3 + \mathrm{A}_5 + \mathrm{A}_7 + \mathrm{A}_9~, 
\]
and similarly for $\mathbf{F}$ in~\eqref{eq:bold-F},
$\mathbf{F} = \mathrm{F}_0 + \mathrm{F}_2 + \mathrm{F}_4 + \mathrm{F}_6 + \mathrm{F}_8 + \mathrm{F}_{10}$. }
\begin{equation}\label{eq:A-basis-definition}
\mathbf{A}\equiv \mathbf{C}\wedge e^{-\mathrm{B}}~.
\end{equation}
For the massless type \Rm{2}A background with $\mathrm{F}_0=0$, the RR field strengths satisfy
\begin{equation}\label{eq:bold-F}
\mathbf{F}=\dd\mathbf{A}\wedge e^{\mathrm{B}}~.
\end{equation}
The six-form component of this relation gives
\begin{equation}\label{eq:dA5-general}
\dd\mathrm{A}_5 = \mathrm{F}_6-\mathrm{B}\wedge \mathrm{F}_4 +\frac{1}{2}\mathrm{B}\wedge\mathrm{B}\wedge \mathrm{F}_2~.
\end{equation}

Using $\mathcal{F}=2\pi\hat{\mathrm f}-P[\mathrm B]$, the classical D4-brane WZ action can be written as
\begin{equation}
I_{\mathrm{WZ}}
=
-T_4\int_{M_5} e^{2\pi \hat{\mathrm{f}}-P[\mathrm{B}]}\wedge P[\mathbf{C}] = -T_4\int_{M_5} e^{2\pi \hat{\mathrm{f}}}\wedge P[\mathbf{A}]~.
\end{equation}
In the gauge adopted in Section~\ref{appx:DBI}, $2\pi\hat{\mathrm f}=0$ on the D4-brane, cf.~\eqref{eq:f-0}. Therefore only the five-form potential contributes,
\begin{equation}\label{eq:IWZ-A5}
I_{\mathrm{WZ}}
=
-T_4\int_{M_5} P[\mathrm{A}_5]~.
\end{equation}

We therefore only need to determine the five-form potential $\mathrm A_5$ for the present background using~\eqref{eq:dA5-general}. We first show that the B-dependent terms in~\eqref{eq:dA5-general} cancel. 
Setting $\mathrm{F}_0 = 0$, $F_{(4)} = 0$ in Eqs.~\eqref{eq:general-B}-\eqref{eq:general-F4-appx}, we now use~\eqref{eq:dA5-general} to obtain $\mathrm{A}_5$. 
It is convenient to introduce 
\begin{equation}
b = \frac{1}{16}e^{2A} \frac{x\sqrt{1 - x^2}}{w}~,\quad \eta = a - \frac{1}{2}y^i g A_{(1)}^i~,\quad K_2 = y^i g F_{(2)}^i~, \quad T_3 = \epsilon^{ijk}g F_{(2)}^i \wedge y^j D y^k~,
\end{equation}
so that we have
\begin{align}
\mathrm{B} &= \ell\, b\, \mathrm{vol}_{S^2} - \frac{\ell}{2}e^A\dd\rho\wedge\eta~,\label{eq:B-s}\\
\mathrm{F}_2 &= -\mathfrak{q}\qty(\mathrm{vol}_{S^2} + K_2)~,\label{eq:F2-s}\\
\mathrm{F}_4 &= -\ell\,\mathfrak{q}\, b K_2\wedge \mathrm{vol}_{S^2} - \ell\frac{\mathfrak{q}}{4} e^A\dd\rho\wedge T_3~.\label{eq:F4-s}
\end{align}
Therefore we have
\begin{equation}
-\mathrm{B}\wedge\mathrm{F}_4 = -\frac{\ell^2}{2}\mathfrak{q}\, b\, e^A\dd\rho\wedge\eta\wedge K_2\wedge \mathrm{vol}_{S^2} + \ell^2\frac{\mathfrak{q}}{4} b\, e^A\mathrm{vol}_{S^2}\wedge\dd\rho\wedge T_3
\end{equation}
and
\begin{equation}
\frac{1}{2}\mathrm{B}\wedge\mathrm{B}\wedge \mathrm{F}_2 = \frac{\ell^2}{2} \mathfrak{q}\, b\, e^A\mathrm{vol}_{S^2}\wedge\dd\rho\wedge \eta\wedge K_2~.
\end{equation}
In our setup, 
\begin{equation}
\begin{split}
\mathrm{vol}_{S^2} &= \sin\theta\dd\theta\wedge\qty(\dd\psi - g A_\tau(r)\dd\tau)~,\\
K_2 &= g A'_{\tau}(r)\cos\theta\, \dd r \wedge \dd\tau~,\\
T_3 &= g A'_{\tau}(r)\sin^2\theta\, \dd r\wedge \dd\tau \wedge \dd\psi~,
\end{split}
\end{equation}
from which one can see that $\mathrm{vol}_{S^2}\wedge T_3 = 0$. 
We then have $\dd\mathrm{A}_5 = \mathrm{F}_6$. 

It remains to determine $\mathrm F_6$ in~\eqref{eq:dA5-general}, which is related to the four-form flux by
\begin{equation}
\mathrm{F}_6 = - *_{10} \mathrm{F}_4~.
\end{equation}
To evaluate the 10d Hodge dual, we write the metric~\eqref{eq:uplift-metric} as
\begin{equation}\label{eq:10d-metric-warped-product-app}
\dd s_{10}^2
=
\mathsf{a}^2 \dd s_7^2
+
\mathsf{b}^2 \dd\rho^2
+
\mathsf{c}^2
\qty[
\dd\theta^2
+
\sin^2\theta\qty(D\psi)^2
]~,
\end{equation}
where we introduce $D\psi = \dd\psi-gA_\tau(r)\dd\tau$ and
\begin{equation}\label{eq:warp-factors-hodge-app}
\mathsf{a}^2
=
\ell\frac{1}{8}g^2 X^{-1/2}e^{2A}~,
\quad
\mathsf{b}^2
=
\ell X^{5/2}~,
\quad
\mathsf{c}^2
=
\ell X^{5/2}
\frac{1 - x^2}{16w}e^{2A}~,
\end{equation}
Although $D\psi$ contains a 7d component, the
fibered metric is diagonal in the adapted orthonormal frame
\begin{equation}\label{eq:adapted-frame-app}
E^a=\mathsf{a} e^a~,
\quad
E^\rho=\mathsf{b}\dd\rho~,
\quad
E^\theta=\mathsf{c}\dd\theta~,
\quad
E^\psi=\mathsf{c}\sin\theta D\psi~,
\end{equation}
where $e^a$ is an orthonormal frame for $\dd s_7^2$. We choose
the orientation
\begin{equation}\label{eq:10d-orientation-app}
\mathrm{vol}_{10}
=
\mathrm{vol}_7\wedge\dd\rho\wedge\mathrm{vol}_{S^2}~.
\end{equation}

For later use, let $\omega_2$ be a two-form along the 7d spacetime and $\lambda_1$ a one-form along the covariantized two-sphere. The adapted frame then gives
\begin{align}
*_{10}\qty(\omega_2\wedge\mathrm{vol}_{S^2})
&=
\mathsf{a}^3\mathsf{b}\mathsf{c}^{-2}
\qty(*_7\omega_2)\wedge\dd\rho~,
\label{eq:hodge-identity-one-app}\\
*_{10}\qty(\dd\rho\wedge\omega_2\wedge\lambda_1)
&=
\mathsf{a}^3\mathsf{b}^{-1}
\qty(*_7\omega_2)\wedge\qty(*_2\lambda_1)~.
\label{eq:hodge-identity-two-app}
\end{align}
Here $*_7$ is defined with respect to $\dd s_7^2$, while $*_2$
is the fiberwise Hodge star associated with
$\dd\theta^2+\sin^2\theta(D\psi)^2$. Using
$\ell=8\sqrt{2}/g^3$, the coefficients simplify to
\begin{equation}\label{eq:hodge-warp-coefficients-app}
\mathsf{a}^3\mathsf{b}\mathsf{c}^{-2}
=
\frac{8e^Aw}{X^2(1-x^2)}~,
\quad
\mathsf{a}^3\mathsf{b}^{-1}
=
\frac{1}{2}e^{3A}X^{-2}~.
\end{equation}

In our setup, $\mathrm{F}_4$~\eqref{eq:general-F4-appx} is given by
\begin{equation}\label{eq:F4-specific-app}
\mathrm{F}_4 = -\ell\frac{\mathfrak{q}}{16}e^{2A}\frac{x\sqrt{1-x^2}}{w}\cos\theta\, g F_{(2)}^3\wedge\mathrm{vol}_{S^2} - \ell\frac{\mathfrak{q}}{4}e^A \dd\rho\wedge g F_{(2)}^3 \wedge\sin^2\theta D\psi~.
\end{equation}
With the orientation in~\eqref{eq:10d-orientation-app}, we have $*_2\qty(\sin^2\theta D\psi) = -\sin\theta\,\dd\theta$. 
Applying~\eqref{eq:hodge-identity-one-app} and \eqref{eq:hodge-identity-two-app} to~\eqref{eq:F4-specific-app},
we obtain
\begin{equation}\label{eq:F6-specific-app}
\mathrm{F}_6 = \ell\frac{\mathfrak{q}}{2} \frac{e^{3A}x}{\sqrt{1-x^2}} \cos\theta\, X^{-2}*_7\qty(g F_{(2)}^3)\wedge\dd\rho - \ell\frac{\mathfrak{q}}{8} e^{4A}\sin\theta\,X^{-2}*_7\qty(g F_{(2)}^3)\wedge\dd\theta~.
\end{equation}
Since the 7d three-form field strength vanishes,
the gauge-field equation~\cite[Eq.~(4.17c)]{Passias:2015gya} reduces to
\begin{equation}\label{eq:Omega5-closed-app}
\dd\qty(X^{-2}*_7\qty(g F_{(2)}^3)) = 0~.
\end{equation}
Moreover, the massless internal BPS equation~\cite[Eq.~(3.5c)]{Passias:2015gya} implies
\begin{equation}\label{eq:e4A-derivative-app}
\dd\qty(e^{4A})
=
4e^{3A}
\frac{x}{\sqrt{1-x^2}}\dd\rho~.
\end{equation}
Eqs.~\eqref{eq:F6-specific-app}-\eqref{eq:e4A-derivative-app}
then show that $\mathrm{F}_6$ is an exact form (here we also use the fact that $\mathfrak{q}$ is a constant)
\begin{equation}\label{eq:A5-solution-app}
\mathrm{F}_6
=
\dd\mathrm{A}_5~,
\quad
\mathrm{A}_5
=
-\ell\frac{\mathfrak{q}}{8}
e^{4A}\cos\theta\,
X^{-2}*_7\qty(g F_{(2)}^3)~.
\end{equation}

For the charged 7d black-hole background, with the orientation $\mathrm{vol}_7 = \dd\tau\wedge\dd r\wedge\mathrm{vol}_{\mathbb{H}^5}$, one finds
\begin{equation}\label{eq:Omega5-black-hole-app}
X^{-2}*_7\qty(g F_{(2)}^3)
=
-gA_\tau'(r)H^2r^5\,
\mathrm{vol}_{\mathbb{H}^5}~.
\end{equation}
Therefore, $\mathrm{A}_5$ is proportional to
$\mathrm{vol}_{\mathbb{H}^5}$. Since the D4-brane wraps only an
$\mathbb{H}^1\subset\mathbb{H}^5$, the pullback of $\mathrm{A}_5$ on the D4-brane vanishes. 
Consequently, the classical WZ action of the probe
D4-brane is zero, $I_{\mathrm{WZ}}=0$. 
\subsection{The D4-brane position}\label{appx:D4-EOM}
We now verify that the embedding $z=j$ satisfies the probe-brane equation of motion in the charged background. The position $z$ should be varied at fixed worldvolume flux~\eqref{eq:flux}.

Keeping $z$ arbitrary in the gauge transformation~\eqref{eq:gauge-transf} gives
\begin{equation}\label{eq:D4-offshell-flux}
\mathcal{F} = \pi\left(j - z + \frac{Z(N-2z)}{Y}\right) \mathrm{vol}_{S^2} + \pi(j - z)y^i gF_{(2)}^i~,
\quad
Y = (N - 2z)^2 + 4X^5Z~.
\end{equation}
Note that the terms proportional to $\dd z$ cancel between $2\pi\hat{\mathrm f}$ and $P[\mathrm B]$. At $z=j$, \eqref{eq:D4-offshell-flux} reduces to~\eqref{eq:flux-new}.

The DBI Lagrangian density at a general constant position $z$ is
\begin{equation}
\mathcal{L}_{\text{DBI}} = 8 \pi^2 k\, r \sin\theta\sqrt{[Z_j+(z-j)^2]^2 + 4 X^5 Z(z - j)^2}\sqrt{1 + \frac{(z - j)^2 A_\tau'(r)^2 \cos^2\theta}{8X^5Z}}~.
    \label{eq:D4-offshell-DBI}
\end{equation}
Both square roots in \eqref{eq:D4-offshell-DBI} have vanishing first derivative with respect to $z$ at $z=j$.
Moreover, for a nonconstant embedding $z=z(\sigma)$, its derivatives enter the induced metric through
\begin{equation}
(g_{\mathrm{D4}})_{ab} = \big(g_{\mathrm{D4}}^{(0)}(z)\big)_{ab} + \frac{2\pi X^{5/2}}{\sqrt{Z}}\partial_a z\,\partial_b z~,
\end{equation}
where $g_{\mathrm{D4}}^{(0)}(z)$ denotes the induced metric at constant $z$.
Since \eqref{eq:D4-offshell-flux} contains no derivatives of $z$, these derivative terms do not contribute to the first variation around $z=j$. Therefore,
\begin{equation}\label{eq:D4-position-EOM}
\left.
    \left[
        \frac{\partial\mathcal L_{\rm DBI}}{\partial z}
        -
        \partial_a
        \frac{\partial\mathcal L_{\rm DBI}}
        {\partial(\partial_a z)}
    \right]
\right|_{z=j}
=0~.
\end{equation}

We next consider the WZ term. For a displacement $V=\delta z\,\partial_z$, Cartan's formula gives
\begin{equation}
    \delta_z P[\mathrm A_5]
    =
    P[\mathcal L_V\mathrm A_5]
    =
    \delta z\,P[\iota_{\partial_z}\dd\mathrm A_5]
    +\dd\left(\delta z\,P[\iota_{\partial_z}\mathrm A_5]\right)~,
\end{equation}
where $\mathcal L_V$ is the Lie derivative along $V$ and $\iota_{\partial_z}$ denotes contraction with $\partial_z$.
The last term is a total derivative.
Moreover, \eqref{eq:A5-solution-app} implies $P[\iota_{\partial_z}\dd\mathrm A_5] = 0$, since this form contains $\mathrm{vol}_{\mathbb H^5}$, whereas the D4-brane wraps only $\mathbb H^1\subset\mathbb H^5$.

At fixed $\hat f$, the transformed worldvolume flux~\eqref{eq:gauge-transf} varies as
\begin{equation}
    \delta(2\pi\hat{\mathrm f})
    =
    \dd\qty[2\pi\,\delta z\,\Big(a-\frac{1}{2}y^i g A_{(1)}^i\Big)]~.
\end{equation}
Using $2\pi\hat{\mathrm f}|_{z=j}=0$ and integrating by parts,
we therefore obtain, up to boundary terms,
\begin{equation}
\left.\delta_z I_{\rm WZ}\right|_{z=j} = -T_4\int_{M_5}\delta z \qty[ P[\iota_{\partial_z} \dd\mathrm{A}_5] + 2\pi\Big(a - \frac{1}{2}y^i g A_{(1)}^i\Big) \wedge P[\dd\mathrm A_3]] = 0~.
\end{equation}
Here the second term vanishes because according to \eqref{eq:bold-F}, the background fluxes satisfy $P[\dd\mathrm A_3] = P[\mathrm F_4-\mathrm B\wedge\mathrm F_2]$, which vanishes given~\eqref{eq:B-s}, \eqref{eq:F2-s}, and~\eqref{eq:F4-s}.
Taking variations that vanish near the boundary removes the boundary terms. Together with the DBI analysis, this verifies the position equation of motion at $z=j$ for general $n$.
\subsection[$\kappa$-symmetry of the D4-brane embedding]{\boldmath $\kappa$-symmetry of the D4-brane embedding\unboldmath}\label{appx:D4-supersymmetry}
We verify that the D4-brane embedding at $z=j$ preserves supersymmetry using the $\kappa$-symmetry condition $\Gamma_\kappa \varepsilon = \varepsilon$, where $\varepsilon$ is a Killing spinor of the 10d background. We perform the analysis in Lorentzian signature.\footnote{We use the analytic continuation $t = -i\tau$, $A_t = iA_\tau$ of the Euclidean saddle. Since $A_t$ is imaginary for $n \neq 1$, the Lorentzian fields and spinors are complexified.} For the D4-brane wrapping $(t,r,\eta,\theta,\psi)$, we choose the induced orthonormal frame
\begin{equation}\label{eq:orthonormal-frame}
\begin{aligned}
e^{\hat{0}}
&=\sqrt{\mathcal{W}_7}H^{-4/5}\sqrt{f}\dd t~,
\quad
e^{\hat{1}}
=\sqrt{\mathcal{W}_7}H^{1/5}\frac{\dd r}{\sqrt f}~,
\quad
e^{\hat{2}}
=\sqrt{\mathcal{W}_7} H^{1/5}r \dd\eta~,
\\
e^{\hat\theta}
&=\sqrt{\mathcal{W}_3}\dd\theta~,
\quad
e^{\hat\psi}
=\sqrt{\mathcal{W}_3}\sin\theta
  (\dd\psi-gA_t\dd t)~.
\end{aligned}
\end{equation}
The gauge-invariant worldvolume flux~\eqref{eq:flux-new} then takes the form
\begin{equation}
    \mathcal{F} = \mathfrak{f}\,e^{\hat\theta}\wedge e^{\hat\psi}~,
    \quad
    \mathfrak{f} = \frac{\varphi}{\mathcal{W}_3}
    =\frac{N - 2j}{2X^{5/2}\sqrt{j(N - j)}}~.
    \label{eq:D4-bps-flux}
\end{equation}
For a D4-brane, the $\kappa$-symmetry operator is~\cite[Eq.~(28)]{Marolf:2003vf}
\begin{equation}
\Gamma_\kappa
=\frac{\epsilon^{\hat{a}\hat{b}\hat{c}\hat{d}\hat{e}}}
{\sqrt{-\det(\eta+\mathcal{F})}}
\qty[
\frac{1}{5!}
 \Gamma_{\hat{a}\hat{b}\hat{c}\hat{d}\hat{e}}\Gamma_{11}
-\frac{1}{2\times 3!}
 \mathcal{F}_{\hat{a}\hat{b}}\Gamma_{\hat{c}\hat{d}\hat{e}}
+\frac{1}{2!\times 2^2}
 \mathcal{F}_{\hat{a}\hat{b}}\mathcal{F}_{\hat{c}\hat{d}}
 \Gamma_{\hat{e}}\Gamma_{11}
]~,
\label{eq:D4-kappa-general}
\end{equation}
where $\hat a,\dots,\hat e$ are tangent-space indices on the D4-brane and we choose $\epsilon^{\hat{0} \hat{1} \hat{2} \hat{\theta} \hat{\psi}} = 1$. In the orthonormal frame~\eqref{eq:orthonormal-frame}, we have $\det(\eta + \mathcal F) = - 1 - \mathfrak{f}^2$ and $\mathcal{F} \wedge \mathcal{F} = 0$. 
The projector~\eqref{eq:D4-kappa-general} therefore reduces to
\begin{equation}\label{eq:projector}
\Gamma_\kappa = \frac{\Gamma_{\hat{0}\hat{1}\hat{2}}}
{\sqrt{1 + \mathfrak{f}^2}}
\qty(
\Gamma_{\hat{\theta}\hat{\psi}}\Gamma_{11} - \mathfrak{f}
)~,
\quad
\Gamma_\kappa^2=1~.
\end{equation}
We use the following Clifford decomposition~\cite[Eq.~(4.24)]{Passias:2015gya}:
\begin{equation}
\Gamma_{\hat{\mu}} = \gamma_{\hat{\mu}} \otimes \mathbf{1} \otimes \tau_2~,
\quad
\Gamma_{\hat{a} + 6} = \mathbf{1} \otimes \sigma_{a} \otimes \tau_1~,
\quad
\Gamma_{11} = \mathbf{1} \otimes \mathbf{1} \otimes \tau_3~.
\end{equation}
Here $\hat\mu=0,\ldots,6$ labels the 7d directions, while $a=1,2,3$ corresponds to $(\hat{\theta},\hat{\psi},\hat{\rho})$.
The Pauli matrices $\sigma_a$ and $\tau_a$ act on the internal spinor and chirality factors, respectively.

The uplifted Killing spinor is $\varepsilon = \varepsilon_1 + \varepsilon_2$, with~\cite[Eq.~(4.22)]{Passias:2015gya}
\begin{equation}
\varepsilon_1 = \qty(\xi^1 \otimes \chi_1 + \xi^2 \otimes \chi_1^c) \otimes \ket{\uparrow}~,
\quad
\varepsilon_2 = \qty(\xi^1 \otimes \chi_2 - \xi^2 \otimes \chi_2^c) \otimes \ket{\downarrow}~.
\label{eq:D4-uplifted-spinors}
\end{equation}
Here $\boldsymbol{\xi} = (\xi^1,\xi^2)^T$ is the complexified 7d spinor doublet, and $c$ denotes internal charge conjugation.
The states $\ket{\uparrow}$ and $\ket{\downarrow}$ have positive and negative 10d chirality, respectively.

The internal spinors are~\cite[Eq.~(4.23)]{Passias:2015gya}
\begin{equation}
    \chi_1 = -i e^{\frac{A}{2}} e^{-i\frac{\pi}{2}\sigma_3} e^{\frac{i}{2}\vartheta\sigma_3} \chi_{\mathbb{S}^2}~,
    \quad
    \chi_2 = e^{\frac{A}{2}} e^{-i\frac{\vartheta}{2}\sigma_3} \chi_{\mathbb{S}^2}~,
\end{equation}
where $A$ is the warp function of the undeformed solution.
The spinor $\chi_{\mathbb{S}^2}$ is a Killing spinor on the unit two-sphere, and the angle $\vartheta$ is defined by $\sin\vartheta=w^{-1/2}x$ and $\cos\vartheta=w^{-1/2}X^{5/2}\sqrt{1-x^2}$.
At $z=j$, \eqref{eq:D4-bps-flux} gives $\mathfrak{f} = -\tan\vartheta_j$, and hence
\begin{equation}
\Gamma_\kappa\qty(\xi^1 \otimes \chi_2 \otimes \ket{\downarrow})
= \qty(\gamma_{\hat{0} \hat{1} \hat{2}}\xi^1)\otimes\qty[-\frac{\sigma_3-i\mathfrak f}{\sqrt{1+\mathfrak f^2}}\chi_2]\otimes \ket{\uparrow} 
= \qty(\gamma_{\hat{0} \hat{1} \hat{2}}\xi^1)\otimes \chi_1\otimes \ket{\uparrow}~.
\label{eq:D4-internal-matching}
\end{equation}
The same matrix maps $\chi_2^c$ to $-\chi_1^c$, accounting for the relative sign in \eqref{eq:D4-uplifted-spinors}.
Since $\Gamma_\kappa$ exchanges chiralities and satisfies $\Gamma_\kappa^2=1$, it suffices to impose $\Gamma_\kappa\varepsilon_2=\varepsilon_1$, which reduces to
\begin{equation}
\Pi_{\mathrm D}\boldsymbol\xi=\boldsymbol\xi~,
\quad
\Pi_{\mathrm D}=\gamma_{\hat0\hat1\hat2}~.
\label{eq:D4-external-projection}
\end{equation}

It remains to exhibit a bulk Killing spinor satisfying~\eqref{eq:D4-external-projection} on the brane.
For the black-hole background, with $g=2\sqrt2$ and in the horizon-regular gauge $A_t=i\sqrt2(H^{-1}-r_H)$, the 7d Killing-spinor equations~\cite[Eqs.~(4.25)-(4.26)]{Passias:2015gya} admit the following solution
\begin{equation}
\boldsymbol{\xi} = r^{1/2} H^{1/10} e^{-\zeta(r) \gamma_{\hat{0}} \rho_3/2} e^{-t\rho_3/(2n)} \chi(y)~,
\label{eq:D4-bulk-spinor}
\end{equation}
where $\rho_3 = \mathrm{diag}(1,-1)$ acts on the 7d doublet and the radial function $\zeta(r)$ is defined by $\cos\zeta = 1/(rH)$ and $\sin\zeta = \sqrt{f}/(rH)$.
The spinor $\chi(y)$ obeys
\begin{equation}
\bar\nabla_{\hat m}\chi=\frac12\gamma_{\hat m}\chi~,
\quad
C\chi=\chi~,
\quad
C=\gamma_{\hat{0}\hat{1}}\rho_3~,
\label{eq:D4-hyperbolic-spinor}
\end{equation}
where $\bar\nabla$ is the spinor covariant derivative on $\mathbb H^5$ and $\hat{m}=2,\ldots,6$.

The commuting involutions $C$ and $\Pi_{\mathrm{D}}$ admit a nonzero initial spinor $\chi_0$ satisfying
\begin{equation}
C\chi_0=\chi_0~,
\quad
\Pi_{\mathrm D}\chi_0=\chi_0~.
\end{equation}
In a frame parallel along the wrapped geodesic, $\chi|_{\mathbb H^1} = e^{\eta\gamma_{\hat{2}}/2} \chi_0$.
Since $\Pi_{\mathrm D}$ commutes with $\gamma_{\hat{0}} \rho_3$, $\rho_3$, and $\gamma_{\hat{2}}$, \eqref{eq:D4-external-projection} holds everywhere on the D4-brane.

Finally, under $t=-i\tau$ and $\gamma_{\hat{0},L} = i\gamma_{\hat{0},E}$, the projection becomes $i\gamma_{\hat{0} \hat{1} \hat{2},E}\boldsymbol\xi = \boldsymbol\xi$.
The Euclidean time factor $e^{i\tau\rho_3/(2n)}$ has the required antiperiodicity under $\tau \to \tau + 2\pi n$ in the polar frame.

\bibliographystyle{JHEP}
\bibliography{biblio}

\end{document}